\documentclass{aa}
\usepackage{graphicx}
\usepackage{color}
\usepackage{amsmath}
\usepackage{natbib}
\bibpunct{(}{)}{;}{a}{}{,}
\usepackage{siunitx}
\usepackage{float}
\usepackage[caption = false]{subfig}
\usepackage{gensymb}
\usepackage{hyperref}
\usepackage{fancyhdr}
\usepackage{multirow}
\usepackage{tablefootnote}
\usepackage{placeins}
\usepackage [english]{babel}
\usepackage{todonotes}
\usepackage [autostyle, english = american]{csquotes}
\MakeOuterQuote{"}

\newcommand{\newtext}[1]{#1} 

\begin{document}

\title{Bridging the Gap: Examining Vision Foundation Models for Optical and Radio Astronomy Applications}

\author{E. Lastufka\inst{1}, 
        O. Bait\inst{3},
        M. Drozdova\inst{1},
        V. Kinakh\inst{1}, 
        D. Piras\inst{1}, 
        M. Audard\inst{2}, 
        M.Dessauges-Zavadsky\inst{2},
        T. Holotyak\inst{1},
        D. Schaerer\inst{2}, 
        S. Voloshynovskiy\inst{1}}

\institute{Department of Computer Science, University of Geneva, 7 route de Drize, 1227 Carouge, Switzerland \email{erica.lastufka@unige.ch}
\and
Department of Astronomy, University of Geneva, 51 Chemin Pegasi, 1290 Versoix, Switzerland
\and
SKA Observatory, Jodrell Bank, Lower Withington, Macclesfield, SK11 9FT, UK}

\titlerunning{Vision Foundation Models for Astronomy}
\authorrunning{E. Lastufka et al.}

\date{Received January 7, 2025; accepted XXXX}

\abstract 
{Vision foundation models, which have demonstrated significant potential in many multimedia applications, are often underutilized in the natural sciences. This is primarily due to mismatches between the nature of domain-specific scientific data and the typical training data used for foundation models, leading to distribution shifts. Scientific data often differ substantially in structure and characteristics, and researchers frequently face the challenge of optimizing model performance with limited labeled data of only a few hundred or thousand images.}
{This work evaluates the performance of vision foundation models in astrophysics, with a focus on identifying the best practices for adapting these models to domain-specific datasets. We aim to establish a framework for selecting, fine-tuning, and optimizing these models for common tasks in optical and radio astronomy.}
{We compared multiple foundation models, including self-supervised, weakly supervised, and distillation-based architectures, across two representative optical and radio datasets. 
Experiments involved different fine-tuning strategies, projector heads, and data preprocessing techniques, with performance evaluated on classification and detection metrics.}
{Features extracted by specific foundation models improved classification accuracy for optical galaxy images compared to conventional supervised training. Similarly, these models achieved equivalent or superior performance in object detection tasks with radio images. However, classification performance for radio galaxy images was generally poor, often falling short of traditional supervised approaches.}
{These findings suggest that selecting suitable vision foundation models for astrophysics applications requires careful consideration of the model characteristics and alignment with the specific requirements of the downstream tasks. This study demonstrates that vision foundation models can be effectively adapted to astrophysical applications, provided practitioners iterate on model selection, training strategies, and data handling. The proposed framework bridges the gap between these advanced models and the unique demands of astronomy, enabling broader adoption of deep learning in the field.}


\maketitle

\section{Introduction}


Large machine learning models, trained on vast amounts of data encompassing multiple domains, have in recent years enabled significant advancements in both language and image processing. Known as foundation models, these serve as a cornerstone for numerous everyday applications. Foundation models are designed either for representation learning, 
in which the goal is to capture essential characteristics of the data, or for generative purposes, where the model attempts to create new data samples similar to the training distribution. In this work, we focus on the potential of the learned representations from vision foundation models applied to astrophysical images.


While machine learning is frequently used to manage large amounts of astrophysics image data, vision foundation models are not yet commonly employed. With the exception of ResNets \citep{he_deep_2016} pre-trained on ImageNet \citep{deng_imagenet_2009}, most works in optical and radio astronomy employ models trained from scratch \citep[e.g.][]{slijepcevic_radio_2024,lochner_astronomaly_2021, becker_cnn_2021}. This includes networks trained for tasks such as image classification and object detection that should be able to benefit directly from the many foundation models available. A more detailed literature review will follow in Sections \ref{sec:classification} and \ref{sec:detection}, but two major themes emerge: firstly, models trained on natural images are not believed to have learned features relevant to astrophysics images, and so are disregarded, and secondly, the desire to employ lightweight models as to not over-fit small labeled datasets precludes the use of modern foundation models, whose smallest architectures tend to have tens of millions of trainable parameters. 

Indeed, the proliferation of foundation models in the recent years has led to a seeming overabundance of choice to the machine learning practitioner. Each model has its strengths and weaknesses; the diversity and complexity inherent in foundation models can pose substantial hurdles for scientists who wish to apply them for domain-specific inquiries. Therefore, the primary goal of this paper is to experimentally address the initial choice of vision foundation model, given properties of astrophysics data, specific downstream tasks (DSTs), and availability of resources for fine-tuning the foundation model.

Vision foundation models fall broadly into three categories, distinguished by their training methods, architectures, and objectives: \textit{self-supervised} models trained on a single modality, \textit{weakly supervised} multimodal models, and \textit{distillation} models that agglomerate learned representations from multiple pre-trained models. 
Supervised models are some of the classic pre-trained models available through software libraries such as \texttt{huggingface}, \texttt{Tensorflow}, and \texttt{PyTorch}; convolutional neural networks (CNNs) trained for image classification like the ResNet family, and more specialized architectures trained for object detection and segmentation like YOLO \citep{redmon_you_2016} and SAM \citep{kirillov_segment_2023}. Self-supervised foundation models do not have a single DST objective, and include Masked AutoEncoders (MAE, \citealp{he_masked_2022}), Masked Siamese Networks (MSN, \citealp{assran_masked_2022}), DINO and its improvement DINOv2 \citep{caron_emerging_2021,oquab_dinov2_2024}. Weakly supervised models enable cross-modal representation learning, first introduced by CLIP \citep{radford_learning_2021}, with its loss function improved upon by SigLIP \citep{zhai_sigmoid_2023}. Knowledge distillation is frequently used to extract smaller models from larger ones, but it can also unify representations of multiple models. AM-RADIO \citep{ranzinger_am-radio_2024} distills a student model from multiple teacher models, allowing it to integrate information from several existing foundation models including CLIP and DINOv2. 

There are also many types of loss functions used to train foundation models, which impact their effectiveness for specialized DSTs. Self-supervised learning frequently uses a contrastive loss to distinguish between different views of the data, or a reconstruction objective to reconstruct parts of an image. Non-contrastive methods, often employing regularization, are aimed at learning stable, robust features without explicit contrastive pairs. Foundation models also employ various backbone architectures with different numbers of parameters, ranging from convolutional networks (CNNs) to Vision Transformers (ViTs, \citet{dosovitskiy_image_2021}). 
Of more concern to scientists is the lack of scientific images in pre-training data. Most foundation models are pre-trained on ImageNet or similar large collections of natural images scraped from the web. Certainly the objectives and specific data used for scientific DSTs are neither known nor used during the training of foundation models. Thus, in accordance with the information bottleneck principle \citep{tishby_information_2000}, it is not obvious whether these models are able to retain features related to the DST, namely sufficient statistics, in their learned representations. If, as this principle suggests, only DST-related information should be kept in the model's representations (and the rest filtered out), it is difficult to determine what, if any, relevant information a foundation model trained on ImageNet might have for astronomy; this is one potential explanation for the aforementioned lack of application of pre-trained foundation models in the literature. Furthermore, there is likely a misalignment between the training data distribution and the DST data distribution; this problem is known in the field of computer science as \textit{distribution shift}. 

In addition to theoretical concerns, practitioners face the challenge of finding \textit{optimal fine-tuning techniques}. Given that labeled data available for scientific applications is typically limited, identifying the appropriate training objective, data augmentations, model architecture, and training hyperparameters is essential to enhance the model's performance.
In this work, we provide a framework to guide the selection of foundation models for astrophysical imaging, tailored to specific characteristics of the target dataset, downstream task, and available resources for fine-tuning.
Through this empirical approach, our study aims to clarify the strengths and weaknesses of vision foundation models, as well as to provide practical insights that can inform model selection and adaptation.




The paper is organized as follows. First we discuss the use of foundation models in astrophysics in Section \ref{sec:useofFM}, along with our choices of illustrative datasets, various foundation models, and downstream task methodology. Section \ref{sec:classification} examines galaxy morphology classification, and several fine-tuning techniques one can use to improve performance on a challenging dataset. A second DST, object detection, is evaluated in Section \ref{sec:detection}. Based on these experiments, we generalize our findings into practical guidelines in Section \ref{sec:bestpractices}, before concluding in Section \ref{sec:conclusion}.

\section{Use of Foundation Models in Astrophysics}\label{sec:useofFM}

\begin{figure*}
\centering
\includegraphics[width=.95\linewidth]{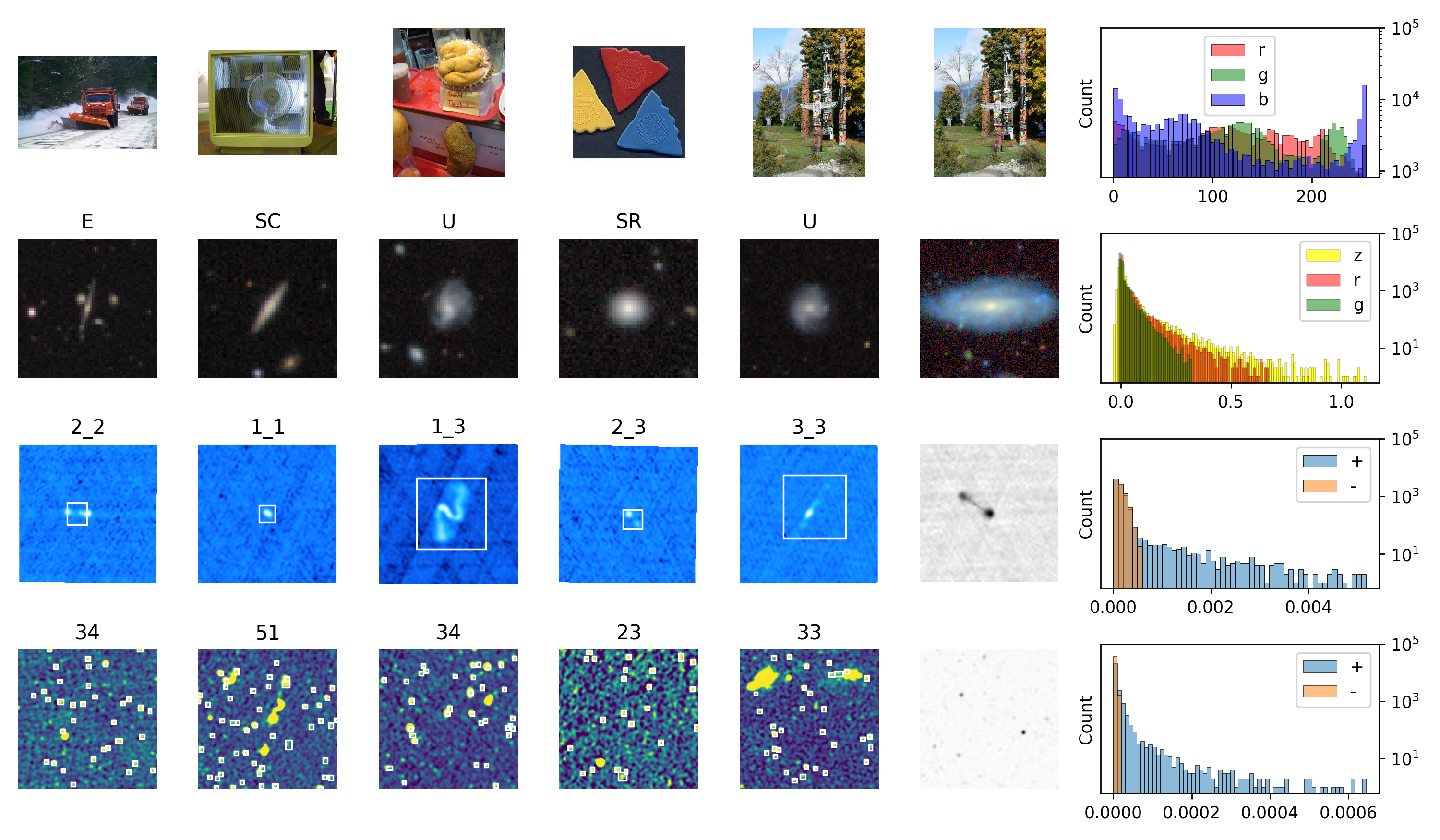}
\caption{\label{fig:sample_data}Random samples from each of the following datasets: ImageNet-1K (top row), GalaxyMNIST (second row), Radio Galaxy Zoo (RGZ, third row), MeerKAT MGCLS (bottom row). GalaxyMNIST and RGZ are labeled according to morphology class, while MGCLS is labeled by the number of compact sources present. White boxes indicate object bounding boxes used for source detection, when applicable. The rightmost two columns display sample images in their raw data format, unscaled, next to the corresponding histogram. GalaxyMNIST combines data from the Dark Energy Camera's \textit{r}, \textit{g}, and \textit{z} channels, while radio images are single-channel continuum images reconstructed from Fourier space, which can therefore include negative values as indidcated by the different colors in the histogram.}
\end{figure*}

Figure \ref{fig:sample_data} illustrates the differences between natural images that comprise common training datasets like ImageNet and images from optical and radio astronomy. Unlike natural images, astrophysics images are found nowhere on Earth and tend to have the following properties: 

\textit{Sparseness}: most of the images consist of several objects that occupy a small fraction of the total image size.

\textit{Noise}: systematic noise is present in the images (this is especially notable for the radio images).

\textit{High dynamic range}: the brightness of the objects in the image can span several orders of magnitude, not easily captured by traditional normalization methods. Research goals most often define the ideal combination of filters or various weighting schemes for displaying the image (for example, to emphasize compact bright sources vs diffuse extended emission).

\textit{Artefacts}: instrumental effects or residuals from image reconstruction can form structures of different scales in the images. 

Because of these fundamental differences, it is \newtext{rare to encounter any deep learning applications in astrophysics literature where networks are trained starting from the pre-trained weights of a foundation model (the only example we found was \citet{burke_deblending_2019}). This is especially true for radio astronomy where images are mathematically reconstructed from sparse samples measured in the Fourier plane.} 

While there are many extremely specialized DSTs in both optical and radio astronomy, often taking advantage of the many wavelength channels available, some more common tasks involving a single or small number of channels are listed in Table \ref{tab:astro_tasks}. Machine learning can offer a number of applicable techniques, and it is possible to use pre-trained foundation models in most cases, \newtext{despite the lack of examples in the literature}.

\begin{table*}[h]
\centering
\resizebox{\textwidth}{!}{
\begin{tabular}{lll}
\hline
Astrophysics Task & Machine Learning Task or Method & Selected References \\
\hline
Image reconstruction & CNNs, de-noising diffusion & \citep{schmidt_deep_2022,drozdova_radio-astronomical_2023} \\
Source detection & Object detection & \citep{vafaeisadr_deepsource_2019,jia_deep_2023, riggi_astronomical_2023} \\
Source characterization & Object segmentation & \citep{farias_mask_2020,sortino_radio_2023} \\
Source classification & Image or object classification & \citep{burke_deblending_2019,riggi_astronomical_2023,merz_detection_2023}\\
Source deblending & Instance segmentation & \citep{burke_deblending_2019,reiman_deblending_2019, hausen_partial-attribution_2022}\\
Object/event discovery & Anomaly detection & \citep{lochner_astronomaly_2021,villar_deep-learning_2021} \\
RFI detection & CNNs, GANs & \citep{vos_generative_2019,li_detection_2021}\\
Background/foreground removal & UNET, de-noising diffusion & \citep{cohen_diffusion-based_2021,zhou_foreground_2023, chen_stability_2024} \\
\hline
\end{tabular}}
\caption{\label{tab:astro_tasks} Common tasks in astrophysics and their machine learning analogues.}
\end{table*}

\subsection{Data}\label{sec:data}

In this work we chose to evaluate galaxy morphology classification and source detection, two tasks common to astronomy across the electromagnetic spectrum. Details of the datasets used to perform these tasks are in Table \ref{tab:datasets}, and sample images are shown in Figure \ref{fig:sample_data}. 

\begin{table*}[h]
\centering
\begin{tabular}{p{1.3cm}p{2.7cm}p{2.8cm}p{5.9cm}p{3.7cm}}
\hline
Dataset & \raggedright{Number of labels} & Image size (pixels) & Data characteristics & Task \\
\hline
GMNIST\tablefootnote{\url{https://github.com/mwalmsley/galaxy_mnist}} & 10K & 64$\times$64 & centered on bright optical galaxy & classification\\
RGZ\tablefootnote{\citep{wong_radio_2024}} & 7.7K & 132$\times$132 & noisy, centered on bright radio galaxy & classification, detection \\
MGCLS\tablefootnote{\url{https://doi.org/10.48479/7epd-w356}} & 10K & 256$\times$256 & many radio sources, varying magnitude & source detection \\\hline
\end{tabular}
\caption{\label{tab:datasets}The datasets used in this study.}
\end{table*}

Classification datasets are images with a single galaxy centered in the cutout. GalaxyMNIST (GMNIST, \citep{walmsley_galaxy_2022}) is a balanced dataset of four categories:  smooth and round (SR), smooth and cigar-shaped (SC), edge-on-disk (E), and unbarred spiral (U). Radio Galaxy Zoo (RGZ) is unbalanced, with labels determined according to number of distinct radio components (C) and number of intensity peaks (P) in each source \citep{wong_radio_2024}. The observed combinations present in RGZ are: 1C 1P, 1C 2P, 1C 3P, 2C 2P, 2C 3P, and 3C 3P, a total of 6 classes. Labeling was done by citizen scientists who performed inspection by varying the relative intensity of continuum radio emission and infrared observations. It is not always visually obvious from the normalized PNG images if there is a difference between a cutout containing the same number of bright peaks -- for example, one containing a single galaxy with two peaks, and another two different components with one peak each.

Source detection datasets also consist of cutouts from wide-field images; they may have one to six galaxies in the central area, as in RGZ, or have tens of galaxies in a single cutout, as in MGCLS. Examples of bounding boxes around sources are shown in Figure \ref{fig:sample_data}; MGCLS labels are consistent in that the boxes only designate compact sources, and not other examples of extended emission or larger sources that might also be present in the images. Because RGZ's labels also contain extended sources, bounding boxes can much larger and filled with a large amount of noise. 

\subsection{Foundation Models}

Foundation models exhibit significant variability in terms of their architectures, training, and performance characteristics. These differences arise from factors such as the size and nature of the training datasets, the number of parameters they incorporate, and their underlying architectural frameworks. Moreover, these models leverage a carefully curated set of data augmentation techniques and diverse pre-training strategies, each tailored to minimize a specific loss function.
Table \ref{tab:models_vk} lists the foundation models investigated in this study. Most were pre-trained on ImageNet-1k's 1.2 million training images. 

\newtext{Some foundation models are pre-trained with a particular downstream task objective, but even if they are not, they can be used as \textit{backbones} together with particular \textit{projector heads} that enable them to be leveraged for a variety of tasks. This involves removing any specialized head that is present for pre-training -- for example, removing the final linear classifier layer of PyTorch's pre-trained ResNet -- and attaching the task-specific head.}

\subsubsection{Categories of foundation models}\label{sec:categories}

Earlier, we broadly categorized vision foundation models into three groups: self-supervised models (SSL), weakly supervised models (WSL), and distilled models. Each model type is characterized by different methodologies and architectures aimed at robust representation learning.

SSL models are designed to extract meaningful representations from large datasets without the need for labels, operating under the assumption that neither the downstream tasks nor the corresponding labels are available during training. This allows SSL models to focus on learning representations that are generalizable and transferable to many different tasks. 
Embeddings can be learned through masking regions or patches of the input image, then inferring the missing content as is done by Masked Siamese Networks (MSN). Embedding reconstruction models, such as Denoising Autoencoders and Masked Autoencoders (MAE), utilize an encoder-decoder architecture to compress the input image into a latent representation and reconstruct it. Alternatively, joint embedding models bypass reconstruction by learning representations for augmented views of the same image, with one encoder fixed as a reference and another trainable encoder processing a masked version. However, these models are susceptible to mode collapse in which embeddings lose meaningful variability. Techniques used by models like SimCLR \citep{chen_simple_2020}, BYOL \citep{grill_bootstrap_2020}, MSN, SwAV \citep{caron_unsupervised_2020}, DINO, Barlow Twins \citep{zbontar_barlow_2021-1}, and VICReg \citep{bardes_vicreg_2022} can mitigate this issue. Hybrid models combining the principles of embedding reconstruction and joint embedding also exist, such as CAE \citep{chen_context_2024} and BeIT \citep{bao_beit_2022}, and a recent innovation is joint embedding prediction models like I-JEPA \citep{assran_self-supervised_2023} and World Models \citep{ha_world_2018}, which introduce an additional projector to predict masked representations to enhance training stability. 

Weakly supervised learning (WSL) models operate in scenarios where full labels are unavailable but auxiliary data, such as text descriptions or metadata, provides partial supervision. These models have dual-encoder architectures, most often with one processing images and another encoding text to generate semantic embeddings.  The objective is to align these embeddings in a shared latent space. 
Representative models in this category include CLIP, GLIP \citep{li_grounded_2022}, SigLIP, CoCa \citep{yu_coca_2022}, and ALIGN \citep{jia_scaling_2021}, which excel in tasks like zero-shot learning and cross-modal retrieval. 

Distilled models rely on a teacher-student paradigm, where a smaller, more efficient model (student) learns to replicate the outputs or distributions generated by a larger, more complex model or models (teacher). The goal is to transfer the teachers' knowledge to the student model; distillation is most often used to create smaller versions of a large foundation model. It can also be used to combine and enhance the outputs of multiple teacher models, such that the student inherits their capabilities across multiple tasks. 
NVIDIA’s AM-RADIO model \citep{ranzinger_am-radio_2024} combines multi-teacher distillation with foundation models, distilling representations from various high-performance models in both standard vision and multi-modal tasks.

\subsubsection{Foundation models used in this study}

\begin{table*}[h]
\resizebox{\textwidth}{!}{%
\begin{tabular}{lllll}
\hline
Model Name           & Backbone       & Number of parameters & Pre-training Dataset   & EMA \\ 
\hline
MAE \citep{he_masked_2022}                 & ViT-Base 16x16 & 86M                  & ImageNet-1k            & -   \\
DINOv2  \citep{oquab_dinov2_2024}             & ViT-Base 14x14 & 86M                  & LVD-142M (proprietary) & +   \\
DINOv1  \citep{caron_emerging_2021}             & ViT-Base 16x16 & 86M                  & ImageNet-1k & +   \\
MSN \citep{assran_masked_2022}                 & ViT-Base 16x16 & 86M                  & ImageNet-1k            & +   \\
SigLIP  \citep{zhai_sigmoid_2023}                & ViT-Base 16x16 & 86M                  & WebLI (English)            &  -  \\
AM-RADIO \citep{ranzinger_am-radio_2024}                & ViT-Base 16x16 & 98.2M                  & various            &   - \\
ResNet 50 \citep{he_deep_2016} & ResNet-50      & 25.6M                & ImageNet-1k            & -   \\
\hline
\end{tabular}
}
\caption{\label{tab:models_vk}The foundation models used in this study.}
\end{table*}

{\textbf{MAE}. The Masked AutoEncoder (MAE) uses masked image modeling (MIM) pretraining, reconstructing masked image patches from only a few visible patches. Unlike natural language processing-inspired approaches like BeIT, which discretize visual tokens through an autoencoder, MAE directly processes the visible image patches through an encoder. The resulting output is then combined with mask tokens to reconstruct the original image using a decoder. Reconstruction error is used as the objective training function.}

\textbf{MSN}. Masked Siamese Networks (MSN) combine MIM with Siamese networks to avoid pixel-level and token-level reconstructions. MSN also uses teacher-student training, with the student network computing the feature representation of a partially masked image.

\textbf{DINO}. DINOv2 improves upon its predecessor, DINOv1 \citep{caron_emerging_2021}, by utilizing a larger and more curated dataset, designated LVD-142M. DINOv2 integrates the DINO cross-entropy loss with the MIM objective employed in iBOT \cite{zhou_image_2022}. 
DINOv2 benefits from the Sinkhorn-Knopp batch normalization technique used in SwAV. 
The model employs a teacher-student training paradigm, where a teacher network computes the feature representation of global views of an image, while a student network computes the feature representation of local views, a series of smaller crops. The  model is optimized to train the student network to replicate the teacher network's output. The teacher network is periodically updated using an exponential moving average (EMA) of the student network's parameters.

\textbf{SigLIP}. Sigmoid language image pre-training (SigLIP), is a multimodal method designed to align image and textual representations in a shared embedding space using a modified contrastive learning framework. 
Pre-trained on WebLI's English text-image pairs, SigLIP uses a sigmoid loss that allows the use of extremely large batch sizes in pre-training, when using multiple GPUs. 

\textbf{AM-RADIO}. 
\newtext{Agglomerative Model – Reduce All Domains Into One (AM-RADIO) merges CLIP, SAM, and DINOv2 into a unified model via multi-teacher distillation. Merging the concepts of a student learning from an ensemble of teachers with foundation models, AM-RADIO trained student models from CLIP, DINOv2, and SAM, the authors used a cosine distance loss to match a simple student adaptor head with the teacher spatial feature vectors and summary vectors (when available). In this way, the student is able to mimic the teacher and perform their targeted downstream tasks.}

In addition to these models, we compare with ResNet, \newtext{one of the most used networks in machine learning for astrophysics}.
Residual Networks are a staple of computer vision, introduced by \citet{he_deep_2016}. They are convolutional neural networks (CNNs) that use skip connections to bypass one or more layers, allowing the network to maintain accuracy over very deep networks. ResNet-18 consists of 8 blocks, while ResNet-50 has 16. 
In this work we use the pre-trained weights available through \texttt{torchvision} \citep{marcel_torchvision_2010}, which were obtained by training in a supervised fashion for classification of ImageNet-1k using cross-entropy loss. A series of very specific data augmentations, including TrivialAugment \citep{muller_trivialaugment_2021}, random erasing, \texttt{mixup} and \texttt{cutmix}, helped increase top-1 accuracy relative to the original augmentation scheme of random resized crops and horizontal flips.

\subsection{UMAP Illustration}

\begin{figure*}[h]
\centering
\subfloat{\includegraphics[width=.161\linewidth,trim={3cm 3cm 4cm 3cm},clip]{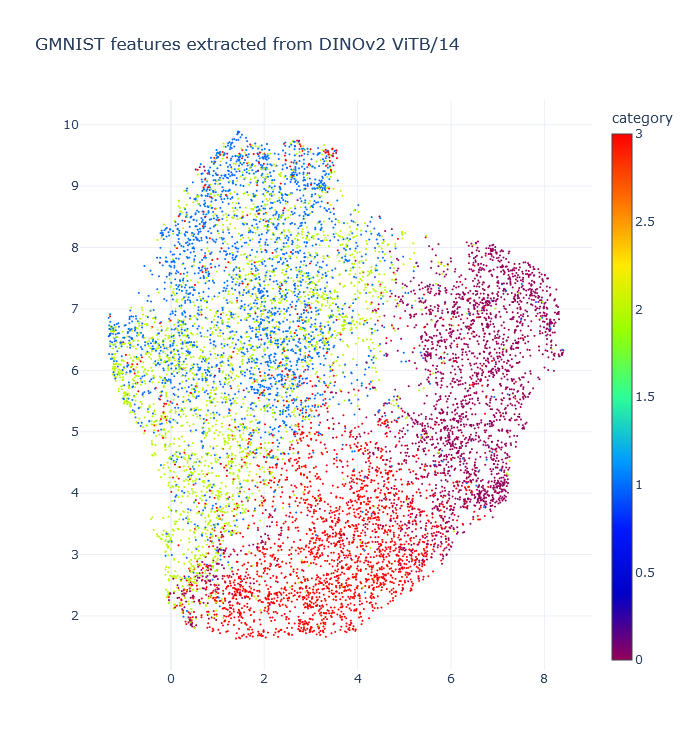}} 
\subfloat{\includegraphics[width=.161\linewidth,trim={3cm 3cm 4cm 3cm},clip]{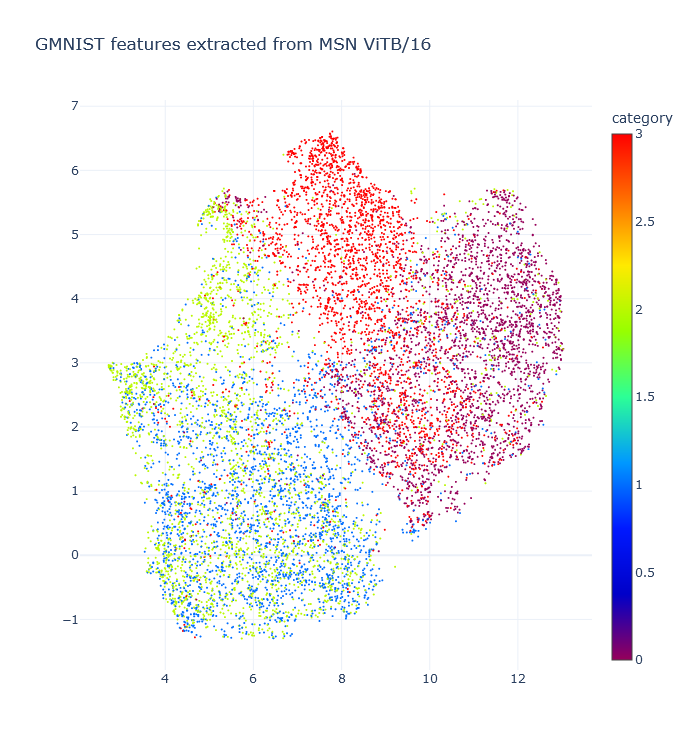}}
\subfloat{\includegraphics[width=.161\linewidth,trim={3cm 3cm 4cm 3cm},clip]{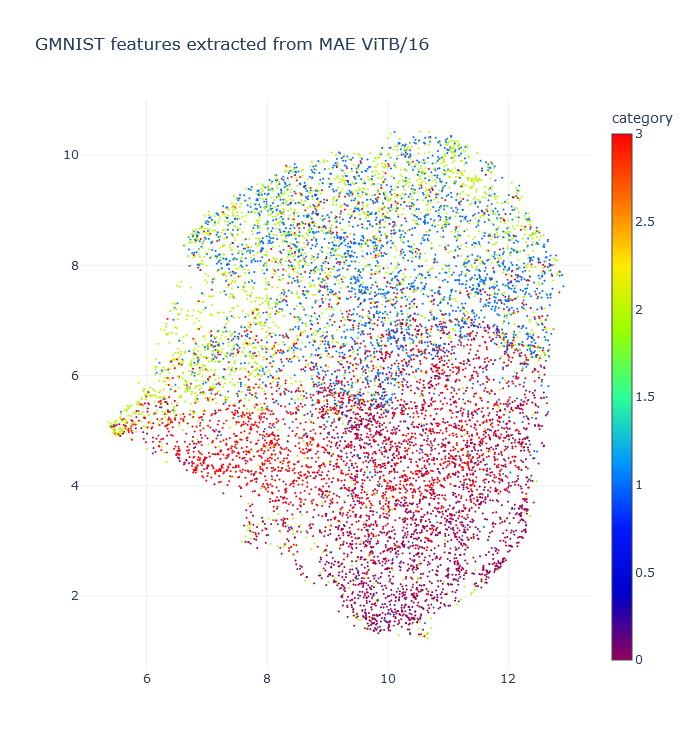}}
\subfloat{\includegraphics[width=.161\linewidth,trim={3cm 3cm 4cm 3cm},clip]{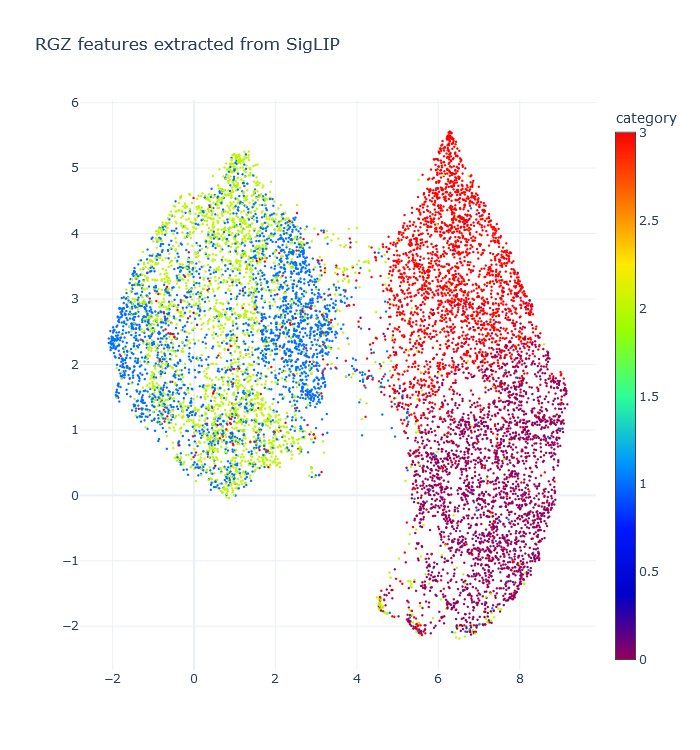}}
\subfloat{\includegraphics[width=.161\linewidth,trim={3cm 3cm 4cm 3cm},clip]{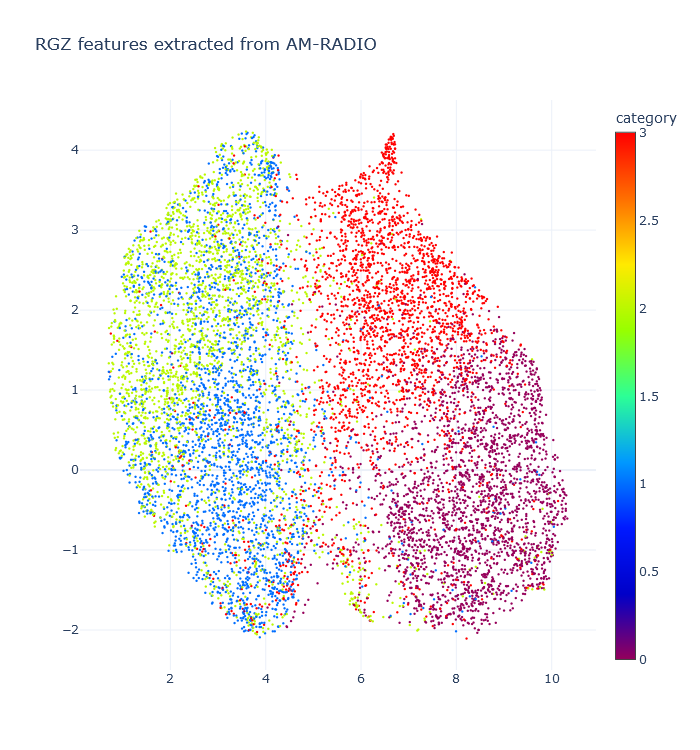}}
\subfloat{\includegraphics[width=.161\linewidth,trim={0cm 0cm 0.5cm 0cm},clip]{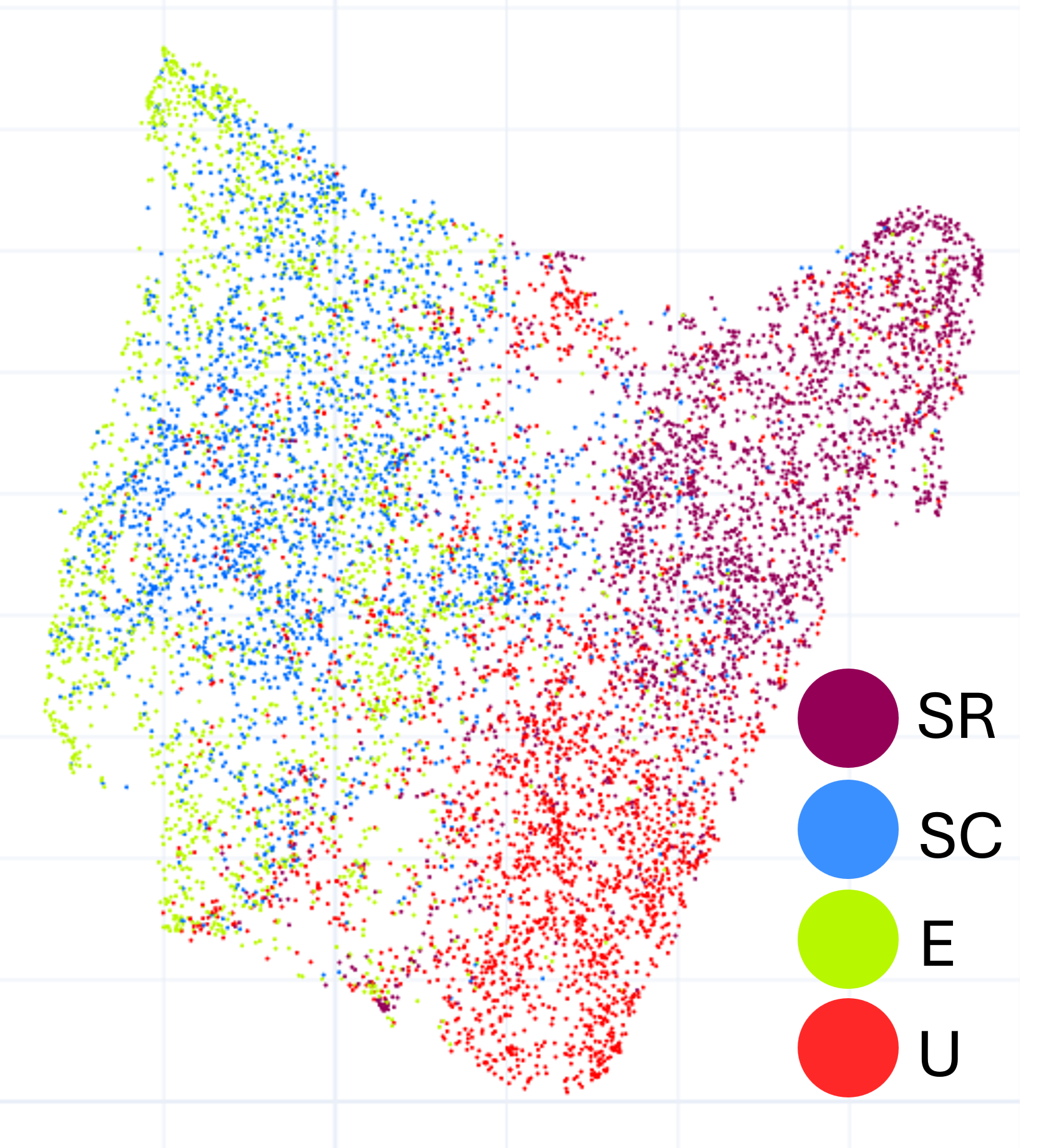}}

\subfloat{\includegraphics[width=.161\linewidth,trim={3cm 3cm 4cm 3cm},clip]{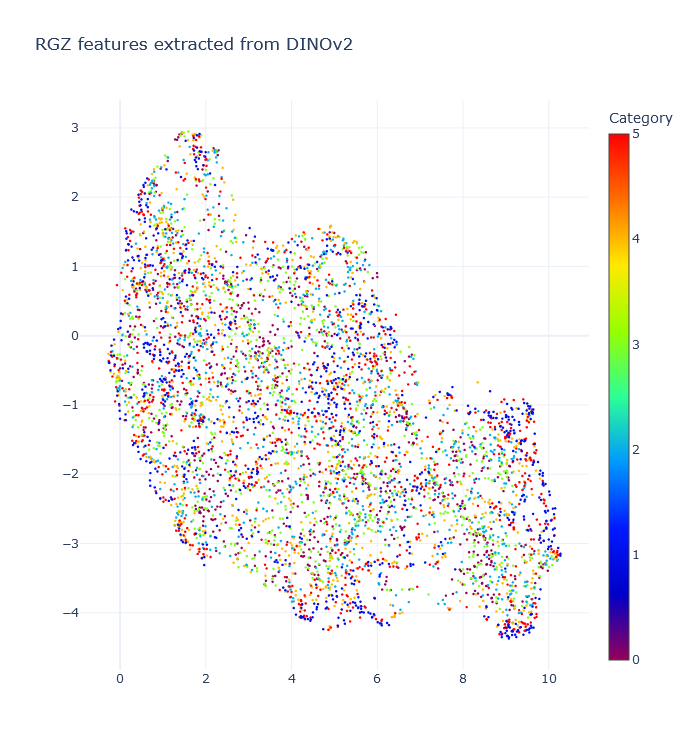}}
\subfloat{\includegraphics[width=.161\linewidth,trim={3cm 3cm 4cm 3cm},clip]{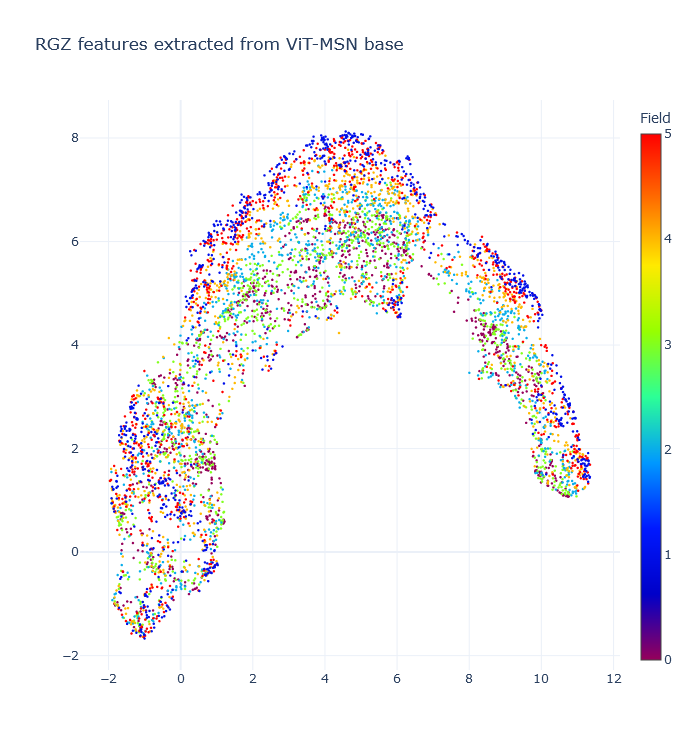}}
\subfloat{\includegraphics[width=.161\linewidth,trim={3cm 3cm 4cm 3cm},clip]{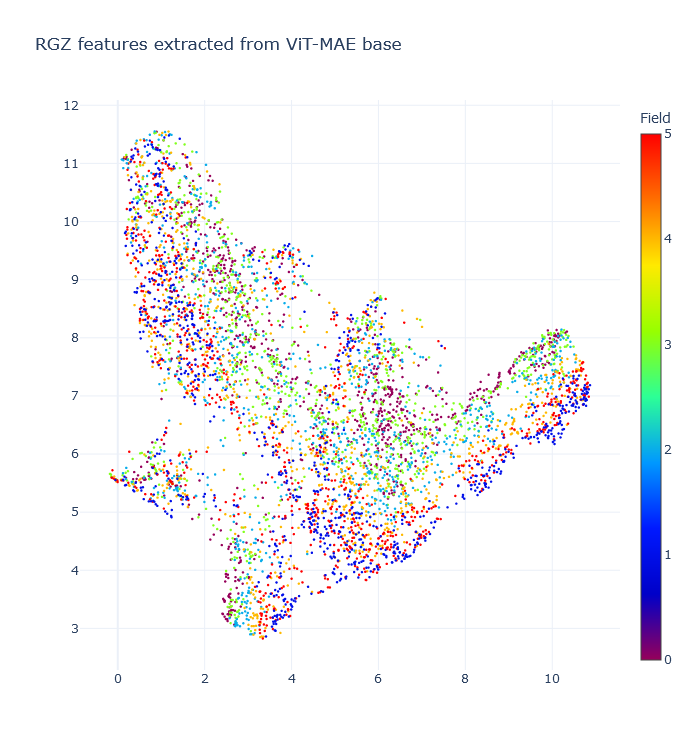}}
\subfloat{\includegraphics[width=.161\linewidth,trim={3cm 3cm 4cm 3cm},clip]{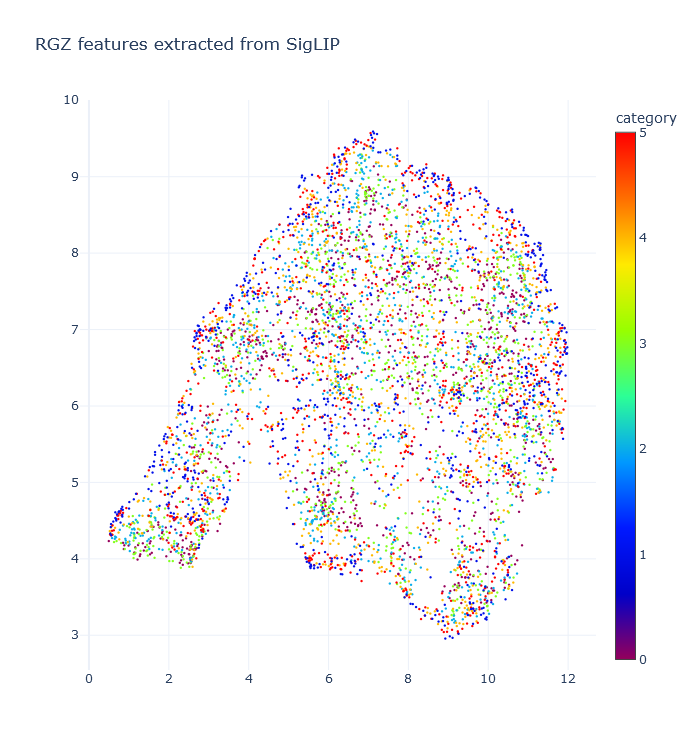}}
\subfloat{\includegraphics[width=.161\linewidth,trim={3cm 3cm 4cm 3cm},clip]{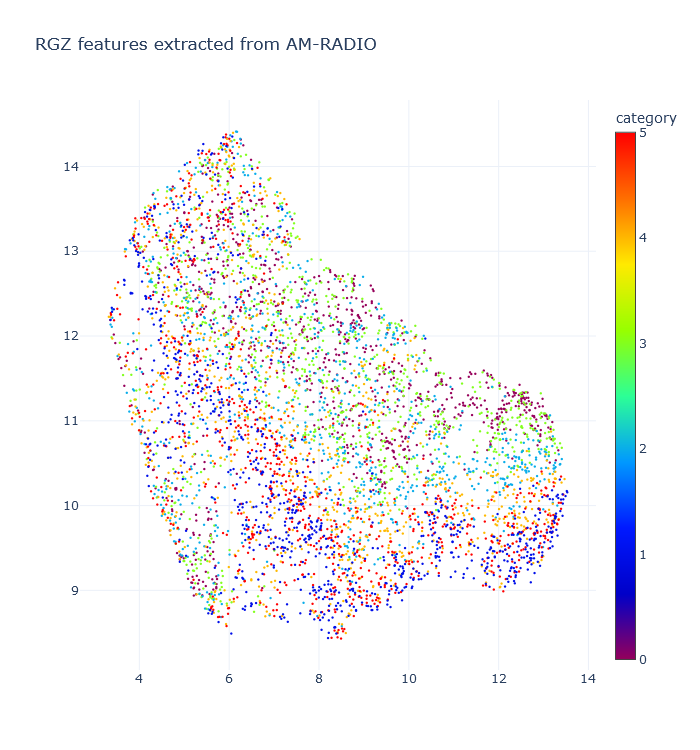}}
\subfloat{\includegraphics[width=.161\linewidth,trim={0cm 0cm 0.5cm 0cm},clip]{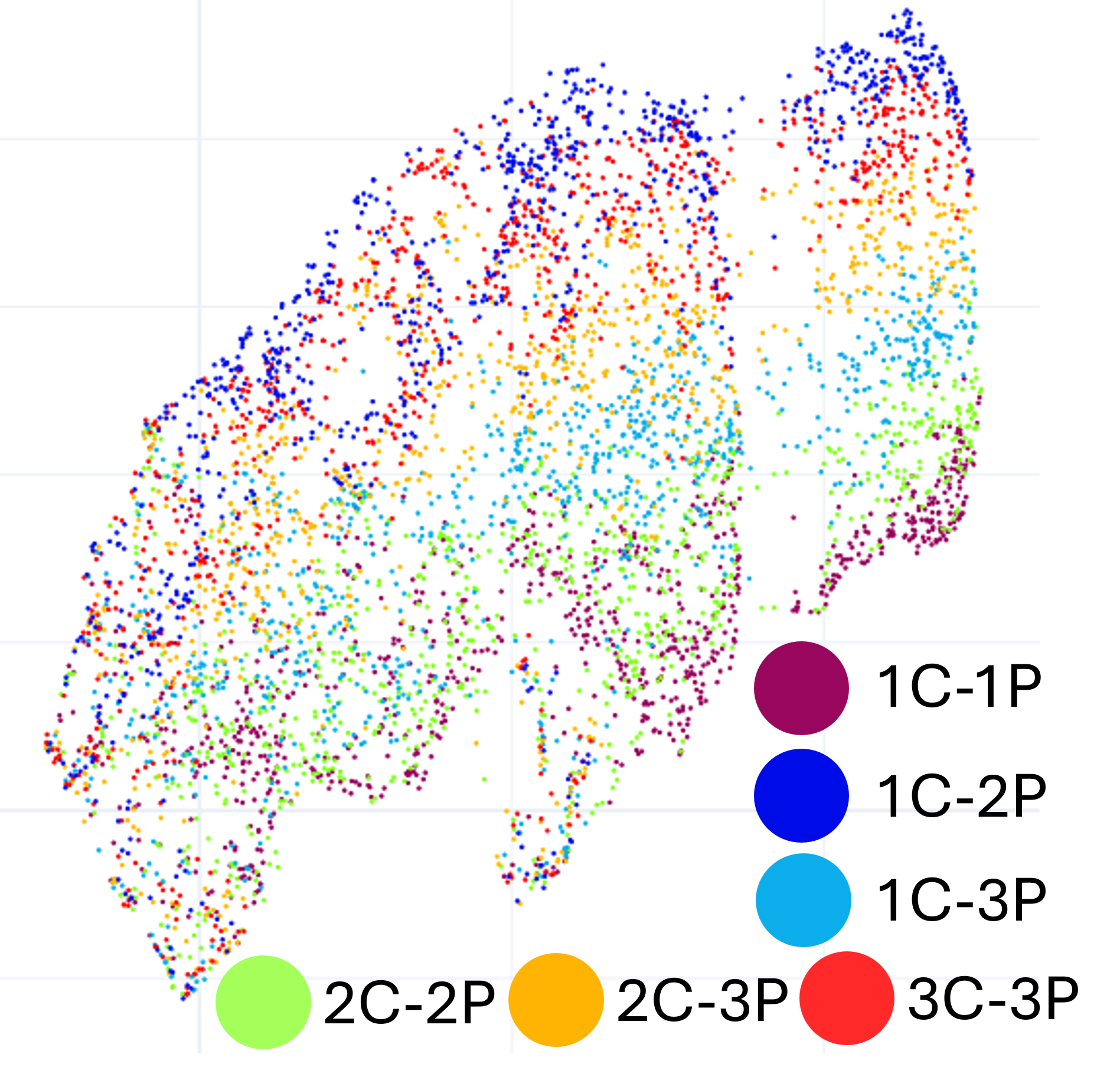}}
\caption{GMNIST (upper row) and RGZ (lower row) UMAP for features extracted using DINOv2, MSN, MAE, SigLIP, AM-RADIO, and ResNet-50 (from left to right). \label{fig:fm_umap}}
\end{figure*}

We provide empirical evidence that foundation models learn different representations of the same data. 
Extracting features from each foundation model and dataset, performing PCA and dimensionality reduction using the first 10 components with UMAP illustrates the latent space (Figure \ref{fig:fm_umap}). We chose UMAP over t-SNE as UMAP better preserves global structure, which is important when local structure in the images might be dominated by noise. For the unbalanced RGZ dataset, a random sample with balanced classes 
is displayed. 

Although UMAP attempts to distill the relationships between thousand-dimensional vectors into two dimensions, it can still reveal clusters where images with similar embeddings reside. This can be deceptive because the model may have not learned relevant embeddings; visualizing the associated class labels can offer insight. 

As shown in Figure \ref{fig:fm_umap}, the UMAP representation of the DINO, MSN, SigLIP, and AM-RADIO feature spaces for GMNIST is quite similar. Classes SR and U, representing smooth-and-round and unbarred spiral galaxies, are in distinct clusters, although MSN features show more overlap between the two than the rest. The majority of SR and U galaxy images are separated by MAE and ResNet-50 as well, although small clusters of these find themselves in other areas. 
Galaxies that appear predominantly elliptical -- smooth-and-cigar-shaped (SC) and edge-on-disk (E) -- share similar latent space embeddings; clustering is most distinct for DINOv2.

Latent space grouping according to class is far less evident with the RGZ data. The best visible separation is seen in ResNet-50 and AM-RADIO. The ResNet model we used was trained with the goal of image classification, as opposed to DINO and MSN, which seek to reconcile local and global image characteristics, and MAE, whose objective is image reconstruction. 

Through visualization of the latent space, we see that our chosen foundation models retain more representations that are class-relevant for GMNIST than RGZ. This is not to say that the learned representations are completely unrelated to the information contained in RGZ, simply that the most important features (according to PCA) are not strongly correlated with image class.



\subsection{Downstream Task Study Methodology}

\newtext{We examine three approaches to finding appropriate models and datasets for tasks. The first is perhaps the simplest: fine-tuning the model to the given dataset. The second, which is often closely related the other two, is to choose the model based on the downstream task. The third is to adapt the data to suit the given model.} 

\newtext{Fine-tuning a foundation model to perform a task on a particular dataset is commonly known as \textit{transfer learning}. This involves unfreezing parameters of the backbone network and allowing those gradients to update. It can be done layer-by-layer or block-by-block, in order to achieve maximum performance with the minimum fine-tuned parameters, or by unfreezing the entire backbone. In this paradigm, the weights and biases of the model itself are optimized for a given dataset and downstream task. In this work, we will consider fine-tuning for both image classification and source detection.}

\newtext{Related to transfer learning is the idea of adapting the model to the downstream task. Both this method and full fine-tuning involve fitting a projector head to the foundation model backbone, to enable it to carry out the DST. However, one can choose to only update the parameters in the head rather than the entire head plus backbone configuration. This significantly reduces computational cost. Choosing a head architecture that is able to filter task-relevant features from the backbone's latent space should perform well for  results for minimal compute.}

\newtext{For this work, the projector head used for \textit{image classification} was a single-layer linear classifier. \textit{Source detection} was done using Faster-RCNN \citep{ren_faster_2017}, whose projector head is implemented slightly different for Vision Transformers and CNNs. For a Vision Transformer backbone, a simple feature pyramid network (FPN) based on only the output of the last large-stride feature map of the backbone is used \citep{li_exploring_2022}. The Faster-RCNN implementation using ResNet also uses a FPN, but one that is heirarchical, acting on feature maps from different convolutional blocks in the ResNet rather than just the last one. After the backbone and FPN, object detection is done by applying a sliding window of the region proposal network (RPN) that predicts whether an object is present or not. These proposals are pooled (RoI pooling) and finally bounding boxes and classes are predicted. We considered the FPN, RPN and RoI layers to comprise the detection head in this situation, as these components are what is attached on top of the backbone network.}

\newtext{The first two approaches are the most familiar to machine learning practitioners; the last will be more intuitive to astrophysicists. Adapting the data to suit the chosen model simply means finding the optimal data pre-processing method, taking into account the characteristics of both the model and the data. In this paper, we closely examine several ways in which this can affect image classification.}


\section{Galaxy morphology classification}\label{sec:classification}

Galaxy morphology classification is a common task in both optical and radio astronomy, and large labeled datasets have been produced thanks to Galaxy Zoo projects \citep{walmsley_galaxy_2022,wong_radio_2024}. Previous deep learning approaches in optical classification have relied heavily on CNNs trained from scratch (\citealp[e.g.][]{zhu_galaxy_2019,maslej-kresnakova_morphological_2021, pandya_e2_2023,urechiatu_improved_2024}). A few works have fine-tuned ImageNet-trained models, such as \cite{hui_galaxy_2022}'s DenseNet121 and \cite{kalvankar_galaxy_2021}'s EfficientNetB5. 
Some works have investigated combining the advantages of ViT's attention mechanism with CNNs, such as \cite{wei_galaxy_2024}, \cite{dagli_astroformer_2023} and \cite{cao_galaxy_2024}. However, application of pure ViTs has been limited and resulted in lower performance than CNNs \citep{lin_galaxy_2022}, although \citet{kumar_vision_2023} reports fine-tuning to be better than training from scratch. Radio galaxy classification work has been more limited due to lack of large  datasets and a well-defined labeling scheme, and so far only employs CNNs; \citet{becker_cnn_2021} and \citet{ndungu_advances_2023} provide comprehensive reviews of the topic. Due to the difficult nature of the data, traditional ML techniques, feature engineering, and finding augmentations specific to radio images, are also active fields of study, while some seek to enhance training datasets with synthetic images using generative deep learning. 

The datasets we use to demonstrate the galaxy morphology classification task, GMNIST and RGZ, are both cutouts of single optical or radio galaxies labeled courtesy of the Galaxy Zoo project. 
Twenty percent of RGZ images contain two labeled galaxies, and four percent contain more than two; these images were all excluded from the classification dataset, reducing the total number of samples to 4692 training images and 1189 test images. Therefore the training data available for RGZ was just over half that available for GMNIST, which was split with 80\% of its 10K images in the training set.

Hyperparameters for the classifier training were similar for both datasets, using a learning rate of 0.0005 and a batch size of 64. Only the classifier head was allowed to train, so the backbone is considered frozen in that the weights and biases of the foundation model backbone stay fixed. Classifiers were trained for 100 epochs on GMNIST and 200 on RGZ. Class weights inversely proportional to the number of samples present in the training dataset were used in the cross-entropy loss function used to classify RGZ. To illustrate the performance differences in low- and high-label regimes, the number of samples used for training was varied at 10\%, 30\%, 50\% and 100\%, while the test set was kept the same. 
For comparison, we trained representative models from scratch in a fully-supervised manner in which both the backbone and classifier head were trained. The fully-supervised models were a ViT-Base model with patch size 16, as well as ResNets with both 50 and 18 layers. The best-performing supervised model was ResNet-50, so its results are shown in Figure \ref{fig:nlabels}.


\subsection{Classification performance}\label{sec:class_perform}

\begin{figure*}[h]
\centering
\subfloat{\includegraphics[width=0.49\linewidth,trim={0cm 0cm 1cm 1.1cm},clip]{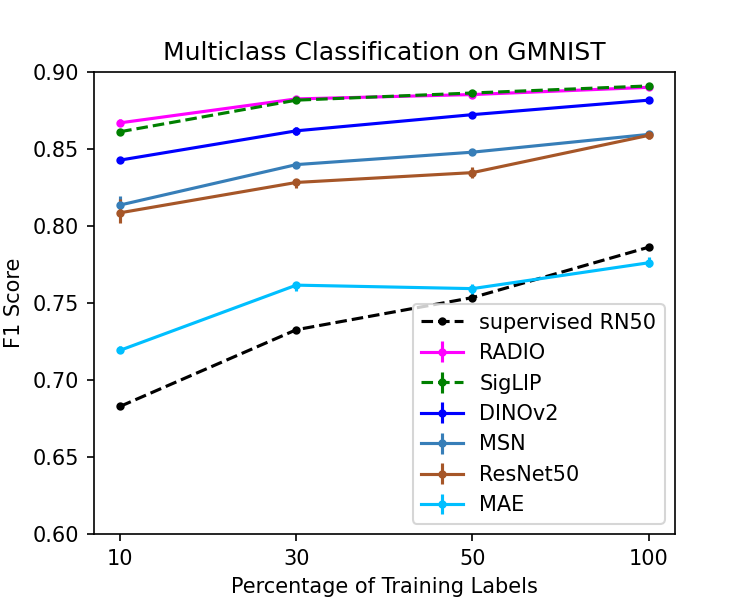}}
\subfloat{\includegraphics[width=0.49\linewidth,trim={0cm 0cm 1cm 1.1cm},clip]{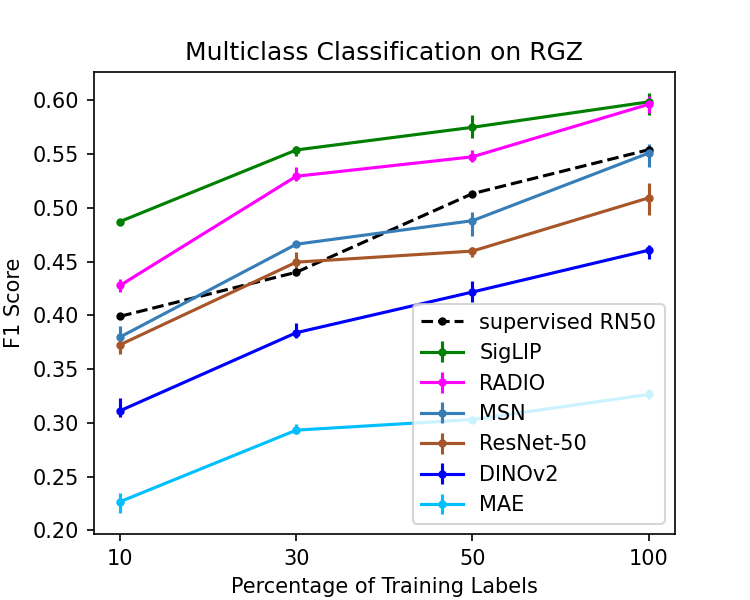}}
\caption{F1 scores for optical (left) and radio (right) galaxy morphology classification, as a function of percentage of training labels. Error bars show the maximum and minimum scores out of three different runs.\label{fig:nlabels}}
\end{figure*}


Figure \ref{fig:nlabels} shows classification F1 score as a function of training label percentage. The F1 score is the harmonic mean of precision and recall, calculated using the number of true and false positives for each class, then averaged. While results differ according to dataset, some trends are in common. Performance generally improved when the number of training labels increased, MAE was the worst performer overall, and the smaller ResNet-18 did not classify as well as ResNet-50. 

On GMNIST, it was notable that the use of a foundation model improved classification relative to fully supervised training by up to 15\%. In fact, the only configuration where starting from a foundation model was outperformed by supervised training is MAE with all available training labels. As supervised training takes hours rather than the tens of seconds required to train a linear classifier layer on frozen, there is a clear benefit to using a foundation model for this particular dataset. Even starting from a model pre-trained on natural images, one observes large increases in performance and speed. 

The distillation model AM-RADIO performed almost exactly as well as SigLIP, with a slight advantage in the low-label regime. Both these models demonstrate a F1 score at least ~0.2 – 0.25 better than the other models. Self-supervised models DINOv2 and MSN both performed better than the ResNets, although ResNet-50 improved the most when trained on 100\% of the data. Unlike DINO and MSN, which use student-teacher paradigms to reconcile partially masked or locally cropped views of images, MAE's objective is to reconstruct an entire image from a few patches. \newtext{\citet{balestriero_how_2024} found that the learned representations with the reconstructive power were the least informative for perception. Our findings support that result; it seems that MAE learns features that are irrelevant to classification}. 

Classification was more difficult with Radio Galaxy Zoo, with F1 scores of less than 0.6 even when using all available training labels. This could be because more of the image is dominated by noise, and less by the galaxy to be classified. The labeling schematic could be another reason for low scores, as discussed in Section \ref{sec:data}. Only SigLIP and AM-RADIO consistently and significantly out-performed supervised training.

The poor performance on radio galaxy classification is not consistent with results from previous works such as \citet{slijepcevic_radio_2024} and \citet{lastufka_self-supervised_2024}, which have shown much higher F1 scores (up to 0.94 for DINOv2) achieved on the MiraBest dataset \citep{porter_mirabest_2023} using pre-trained foundation models. MiraBest is designed for binary classification in the Fanarhoff-Riley scheme, which could be roughly described as a single Gaussian component with one peak versus a single component with two peaks. 

\begin{figure*}[h]
\centering
\includegraphics[width=0.49\linewidth,trim={0cm 0cm 1cm 1.1cm},clip]{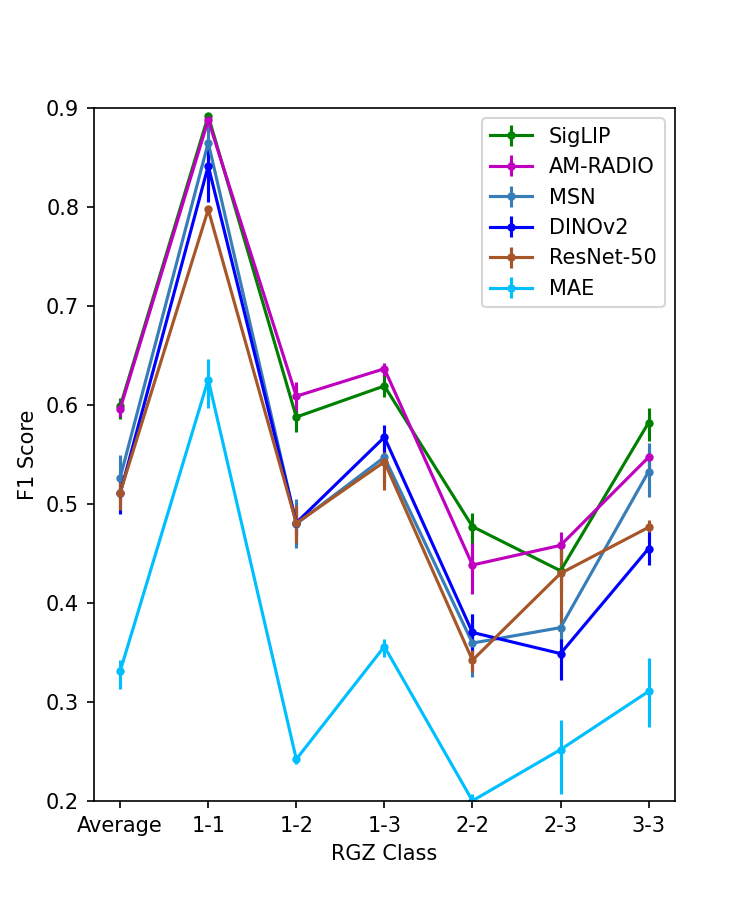}
\includegraphics[width=0.49\linewidth,trim={0cm 0cm 1cm 1.1cm},clip]{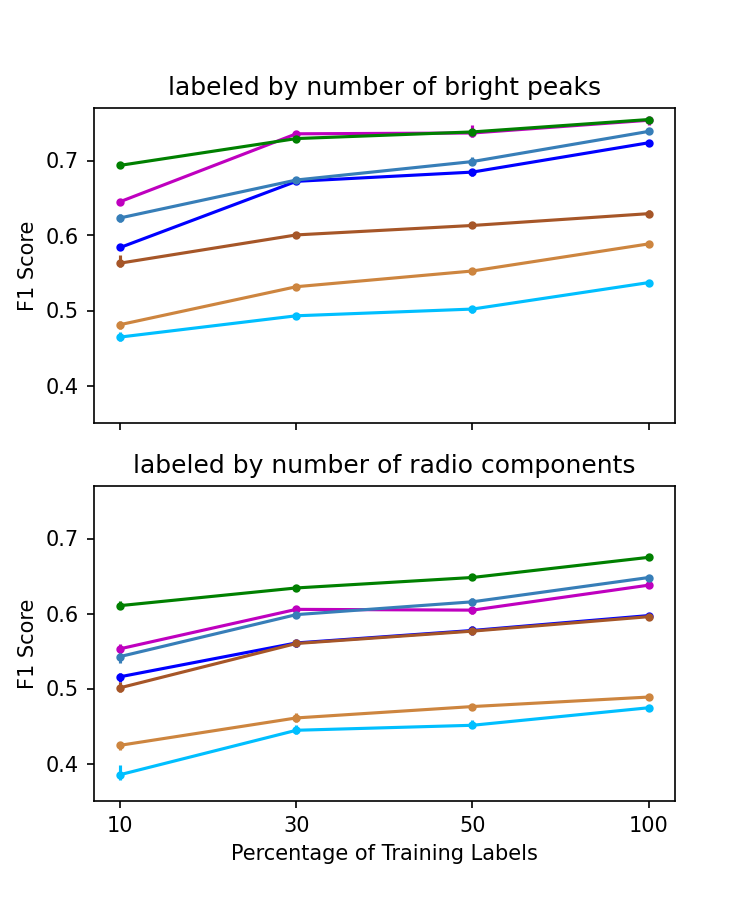}
\caption{Left: F1 score per class for classification on RGZ. Models can confidently identify single Gaussian sources but struggle with multi-peaked or multi-component sources. Right: Label scheme is re-defined to the number of bright peaks (top) or number of individual radio components (bottom). Higher F1 scores in the top chart show that vision models are better at distinguishing bright peaks in images rather than flux islands defining radio components.\label{fig:relabel}}
\end{figure*}

 Challenges with the RGZ dataset may stem from the more subtle characteristics that distinguish the classes. Figure \ref{fig:relabel} (left side) shows that classification for galaxies with two radio components and either two or three bright peaks is especially poor. Re-labeling the images by the number of bright peaks resulted in F1 scores of up to 0.74 (seen on the right side of Figure \ref{fig:relabel}), with all models but MAE out-performing a supervised baseline. Conversely, when re-labeling the images by the number of distinct radio components, the maximum F1 score achieved was 0.63, from MSN with all labels. 

It seems that foundation models struggle to identify distinct radio sources, especially when they involve multiple emission peaks. 
Radio flux islands, containing one or more peaks, are usually identified through analytic source-finding algorithms via connected-component labeling, starting by selecting regions of pixels above a certain flux threshold. This threshold is usually up to 10 times the background RMS, and the brightest pixels it contains can be orders of magnitude greater. This information of relative flux is lost after compression and normalization of the images. 

Unlike GMNIST, RGZ images are reconstructions which contain noise with strong inter-pixel correlations on the scale of the synthesized beam.
These statistical differences between ImageNet and radio images suggest that a well-performing model may require knowledge of a very specific set of features; a true case of distribution shift.

There are two immediately recognizable visual differences between GMNIST, on which classification worked well, and RGZ (see again Figure \ref{fig:sample_data}): radio galaxies occupy a much smaller area of the total image than optical ones, and background noise is much higher and patterned. \newtext{We experimented with bringing the RGZ dataset into better alignment with GMNIST by leveraging the bounding box information to make very close crops around each radio galaxy. This resulted in a F1 score of 0.70 for all labels, using AM-RADIO to extract frozen features.} 

\begin{figure}[h]
\centering
\subfloat{\includegraphics[width=0.49\linewidth]{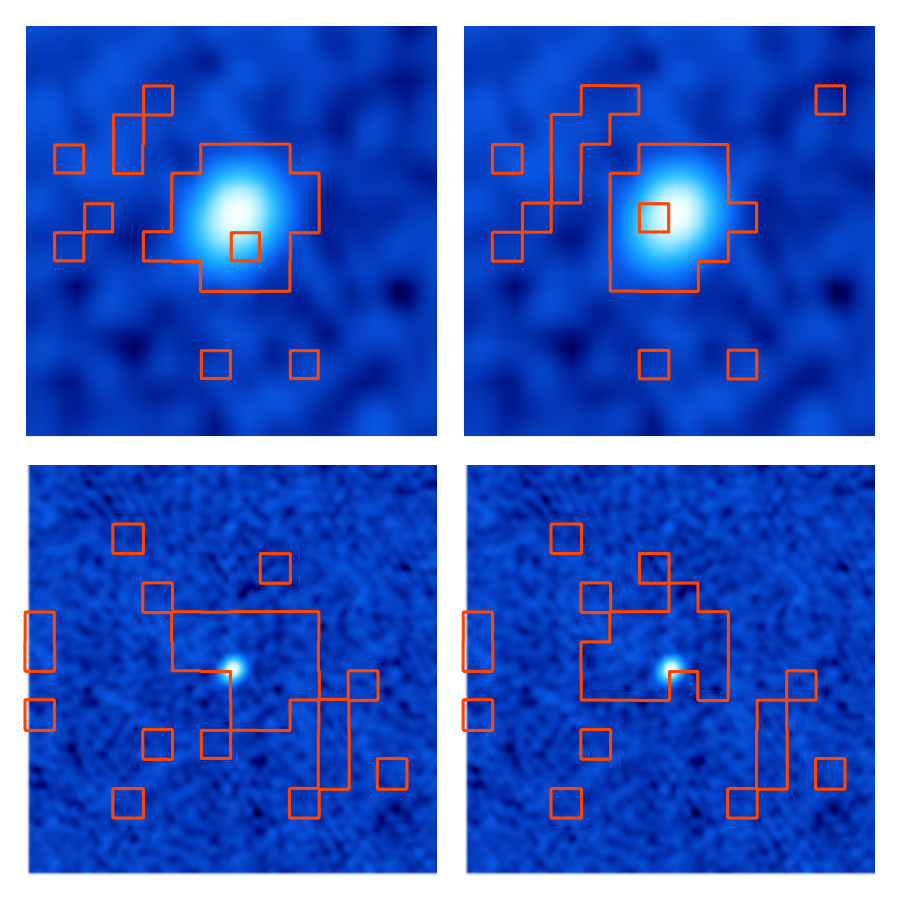}}
\subfloat{\includegraphics[width=0.49\linewidth]{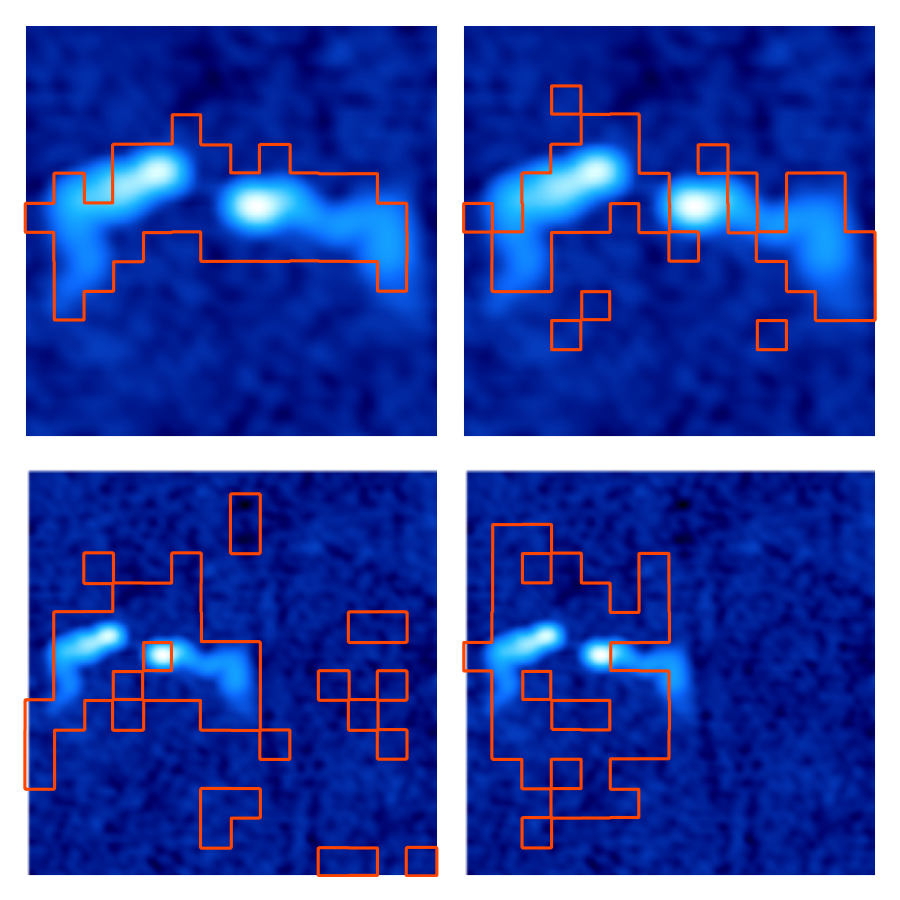}}
\caption{Attention maps (red contours) overlaid on closely-cropped (top) and uncropped (bottom) RGZ images. A typical compact source (labeled as 1C-1P) is shown on the left, with a more complex source with diffuse elements (labeled as 2C-3P) on the right. The attention maps are shown for two different attention heads as contours at the 90th percentile level. \label{fig:attention}}
\end{figure}

\newtext{While this is not yet on par with classification performance on GMNIST, it is clear that the foundation models are able to extract more information from the closely cropped images. Figure \ref{fig:attention} shows the output of selected attention heads on both the cropped (top row) and uncropped (bottom row) RGZ images for two different classes of galaxy. Even though the 1C-1P class of the left side image is in general confidently classified, the attention maps show that areas of high interest to the different attention heads correspond mostly to noise in the uncropped images. The central compact source of the left image only occupies about one 16$\times$16 pixel patch in the uncropped image, which has already been upscaled from its original size.} Attention maps for sample images from all datasets and all transformer-based backbones can be found in Appendix  \ref{sec:attention}.
While the out-of-the-box application of foundation models to GMNIST images performed well, it is not sufficient for classification on RGZ where it offers little to no advantage over supervised training, without careful adaptation to suit the chosen foundation model.

\subsection{Optimization strategies}\label{sec:finetuning}


We demonstrate various common strategies for improving classification performance on a challenging dataset like RGZ. These include, roughly in order of the time it takes to implement the changes:
\begin{itemize}
    \item Use a multi-layer perceptron (MLP) for classification instead of a single linear layer.
    \item Use a bigger model with more parameters.
    \item Use different pre-processing methods.
    \item Fine-tune the backbone via Low-Rank Adaptation (LoRA).
    \item Fine-tune the backbone using a semi-supervised method.
\end{itemize}

\newtext{The first two strategies consist of adapting the model, the third adapts the data to suit the model, and the last two are different ways to implement fine-tuning.}
For purposes of illustration, we report the results using all available training labels and the models AM-RADIO and ResNet-50 in Table \ref{tab:fine-tuning}. With the notable exceptions of employing a close crop and fine-tuning, these techniques improved the classification result of AM-RADIO more than that of ResNet-50. In fact, using either larger input image sizes or larger networks decreased the accuracy of ResNet. 

The effects of each technique is reported on its own; one can expect that employing multiple techniques is a little less than the sum of each individual improvement. For example, using AM-RADIO and combining a close crop and fine-tuning the backbone gives an F1 of 0.82, rather than 0.84. We briefly describe each technique and its impact on classification score below.

\begin{table}[h]
\centering
\begin{tabular}{lrr}
\hline
           Technique &  ResNet-50 F1 &  AM-RADIO F1 \\
\hline
Whitening Layer &    0.519 (+0.008) &     0.603 (+0.013) \\
            MLP &    0.526 (+0.015) &     0.623 (+0.034) \\
RN101/ViT-Large &    0.486 (-0.025) &     0.625 (+0.036) \\
 RN152/ViT-Huge &    0.486 (-0.026) &     0.653 (+0.063) \\
           Crop &    0.643 (+0.132) &     0.705 (+0.115) \\
   Upsample 320 &    0.481 (-0.031) &     0.630 (+0.041) \\
   Upsample 480 &    0.465 (-0.046) &     0.666 (+0.080) \\
   Upsample 512 &    0.454 (-0.057) &     0.667 (+0.078) \\
           LoRA &    0.674 (+0.163) &     0.692 (+0.102) \\
    Fine-tuning &    \textbf{0.737 (+0.226)} &     \textbf{0.721 (+0.132)} \\\hline
\end{tabular}
\caption{Effect of various optimization techniques on RGZ classification. 
}\label{tab:fine-tuning}
\end{table}


\subsubsection{Multi-layer perceptron}

So far, classification was done using a single linear layer as a classifier. Increasing the number of fully-connected layers in the classifier head lends a few advantages. Mainly, the addition of at least one hidden layer enables the classifier to learn non-linear functions. These dense networks with one or more hidden layers are commonly referred to as MLPs (multi-layered perceptrons). 

For a two-layer MLP, we observe a 5.7\% increase in the F1 score of AM-RADIO and a 2.9\% increase for ResNet-50.

We also investigate the use of a whitening layer \citep{kalapos_whitening_2024}, which scales the frozen features before input to the linear classifier layer. Effectively, this is a two-layer MLP with the weights and biases of the input layer pre-calculated. This leads to a 2.2\% increase in the F1 score of AM-RADIO and a 1.5\% increase for ResNet-50.

\subsubsection{Bigger model}

In general, performance increases with the number of parameters in a model \citep[e.g.][]{zhai_scaling_2022}. AM-RADIO has pre-trained two models larger than 86M parameter ViT-Base: ViT-Large with 307M parameters, and ViT-Huge with 632M parameters. Larger ResNet variants are ResNet-101, with 42.8M parameters, and ResNet-152, with 58.5M parameters, all pre-trained for ImageNet classification.

Using features extracted from both these models, we observe an increase in the F1 score of AM-RADIO of 8.1\% (ViT-Large) and 10.7\% (ViT-Large). ResNets with more parameters did not lead to an improvement, but rather the opposite; the F1 score decreased by 5\% when using features extracted from ResNet-101, and by about the same amount for ResNet-152. 

\subsubsection{Custom data pre-processing}

Like most scientific images, optical and radio astronomy data is not delivered straight from the telescopes. Raw data must be calibrated, to account for instrumental or environmental effects as much as possible, and sometimes further processed like in the case of imaging the visibility data collected by radio arrays. Much more so than the natural image domain, the data reduction pipeline has a large impact on the capabilities of the resulting dataset.

In the case of RGZ images, a number of steps have taken place before the final images are incorporated into the dataset. First, radio sources were identified from a survey catalog, which is itself the result of a source-finding algorithm. Next, various filters such as signal-to-noise ratio determined if the image source was kept in the sample or not. Were archival images not available, the initial image would have had to be reconstructed. Experts then had to decide on a cutout size and how to scale each image, which consist of float values spanning five or more orders of magnitude, 
 before finally converting the image to a PNG. Often, both image reconstruction and scaling favor a certain type of radio emission over another, making a dataset more suited to certain scientific goals or downstream tasks than others.

The final product of most astrophysics image reduction pipelines are 16- to 32-bit FITS images in physical units, rather than 8-bit PNG scaled between 0 and 255. As only PNG images were available in the GMNIST and RGZ datasets, we can only examine pre-processing methods that operate on those images, which have already lost some detail. 
One way in which the data has been adapted to fit the model was already mentioned previously; the images had to be re-sized in order to be accepted as input to most models with a ViT-Base backbone. CNNs do not have this input size requirement of 224$\times$224 pixels, as the convolution kernel can operate on images of any size and shape. Similarly, AM-RADIO was designed to take any input image size as long as the dimensions are a multiple of 14.

We resize the RGZ images to 320, 480, and 512 pixels square using bilinear interpolation. Interestingly, increasing the image size does not necessarily increase performance; the optimal size for AM-RADIO is 480$\times$480, leading to a F1 score increase of 8\%. As with increasing the model size, increasing image size has a detrimental effect on the performance of classification via ResNet-50.

Another pre-processing method already mentioned is making a close cutout of the radio source based on the bounding box information. Cropping to the area around the central galaxy leads to a large improvement in F1 score: 16\% increase for AM-RADIO, and 26\% increase for ResNet-50. 
Similar pre-processing, involving cropping images to less than 50\% their original area, was used since \citet{dieleman_rotation-invariant_2015} and \citet{zhu_galaxy_2019} to classify optical galaxy images from SDSS and DeCALs dataset; more recent works like \citet{wei_galaxy_2024} and \citet{urechiatu_improved_2024} also use this method to achieve F1 scores above 0.95. 
As seen from the attention maps in Figure \ref{fig:attention}, the close crop means that more patches of the ViT contain parts of the object to be classified, allowing the network to use information from more than just one or two patches to perform classification. Custom pre-processing allows the data to complement features the foundation model has already learned, and emphasizes the importance of domain-specific expert knowledge in the development of machine learning pipelines.


\subsubsection{Full fine-tuning of the backbone model}

This is traditional fine-tuning: unfreezing parameters the backbone network and allowing them to update. It can be done layer-by-layer or block-by-block, but for this experiment we simply unfreeze the entire backbone after ten epochs, which are used to allow the newly initialized classifier layer to acclimate. 

When fine-tuning the backbone, we observe an increase in the F1 score of AM-RADIO of 22\% and 44\% for ResNet-50. This is the largest improvement in performance by any of the optimization techniques we investigated.


\subsubsection{Fine-tuning the backbone model using LoRA}

Low-Rank Adaptation, or LoRA, is a parameter-efficient fine-tuning method. Instead of directly updating model weights, LoRA tracks changes to the weights and then decomposes the large matrix of weight changes into smaller matrices. These low-rank matrices are multiplied to approximate the full matrix of weight changes, allowing for fast and efficient fine-tuning. In practice, it is not much more difficult to implement than full fine-tuning, although there are additional hyperparameters to be optimized. 

Using LoRA leads to a 17.3\% increase in the F1-score with AM-RADIO as a backbone, and a 31.8\% increase when using ResNet-50. These are substantial improvements that result from tuning less than 5\% of the model parameters.

\section{Source detection}\label{sec:detection}

Source detection in both optical and radio images is largely performed analytically. Work using deep learning to replace these time-consuming analytic pipelines has been done by \citet{wu_radio_2019, farias_mask_2020, burke_deblending_2019, riggi_astronomical_2023}, mainly through application of Faster- or Mask-RCNN. \citet{sortino_radio_2023} provides an overview of both CNN and ViT-based methods, including YOLO. More recent results based on simulated radio data include \citet{taran_challenging_2023} and \citet{cornu_yolo-cianna_2024}, which both surpass the capabilities of analytic source detection.

Here we performed source detection 
on radio continuum datasets RGZ, which mostly contains single galaxies labeled by morphology, and MGCLS, where a single image contains from 10-100 individual compact sources. MGCLS images may also contain larger sources which are not labeled. 

\begin{table*}[h]
\centering
\begin{tabular}{lll|ll}
\hline
Backbone & RGZ mAP@50 & & MGCLS mAP@50 & \\
 & frozen & fine-tuned & frozen & fine-tuned  \\ 
 \hline
ViT-Base supervised & - & 0.51 & - & 0.885 \\
MSN & \textbf{0.70 (+0.19)} & 0.71 (+0.20) & 0.878 (-0.006) & 0.891 (+0.007) \\
DINOv1 & 0.63 (+0.12) & 0.73 (+0.22) & \textbf{0.886 (+0.001)} & \textbf{0.896 (+0.011)}\\
MAE & 0.51 (+0.00) & 0.73 (+0.22) & 0.867 (-0.018) & 0.873 (+0.005)\\ 
SigLIP & 0.46 (-0.05) & 0.679 (+0.17) & 0.872 (-0.013) & 0.878 (-0.007) \\ 
AM-RADIO & 0.47 ( -0.04) & 0.605 (+0.09) & 0.867 (-0.018) & 0.885 \\ 
\hline
ResNet 50 supervised & - & 0.723 & - & 0.873 \\
ResNet 50  & 0.627 (-0.096) & \textbf{0.778 (+0.05)} & 0.849 (-0.024) & 0.873 (+0.00) \\ \hline
ResNet 18 supervised & - & 0.644 & - & 0.511 \\
ResNet 18  & 0.311 (-0.333) & 0.646 (+0.002) & 0.020 (-0.490) & 0.511 (+0.00) \\
\hline
\end{tabular}
\caption{\label{tab:RGZsource_detection} Source detection mAP@50 for the RGZ and MGCLS datasets. Results for frozen and fine-tuned backbones are compared with supervised baselines.}
\end{table*}

As was done for classification, images were upscaled from their native resolution. For RGZ, scaling to 224$\times$224 meant that the average single object size became 55$\times$55 pixels, while the 25th percentile was 24$\times$24 pixels. MGCLS crops were doubled in size from 256$\times$256. This was done so that the average compact source occupied 16$\times$16 pixels, and the 25th percentile 14$\times$14 pixels. The size of the sources relative to the transformer patch size is important; sources unable to occupy most of a patch were difficult to detect. This was proven empirically -- performance on detection with MGCLS increased by more than 10\% when the image size doubled. 
Otherwise, we kept the default training data augmentations: a randomly applied blur, contrast adjustment, and color jitter. 

Our first experiment kept the backbones frozen, allowing only Faster-RCNN's detection head layers (FPN, RPN and ROI) to train. Due to differences in architecture this resulted in a different number of trainable parameters: 21M for ViT-Det, which was used for all the ViT-based models, 17.8M for ResNet-50 with FPN, and 11M for ResNet-18. Because of the structure of the FPN for ViT-Det, for this task the DINO backbone used was DINOv1, which had available pre-trained weights for the architecture ViT-Base with 16x16 patch size (DINOv2 uses 14x14). In the second experiment we allowed the weights and biases of the backbone networks to update as well, in the process known as fine-tuning.

Generally, a learning rate of 0.001 was used, with the exception of training ViT-Base on MGCLS from scratch, and both frozen and fine-tuned MAE on MGCLS, where we used a learning rate of 0.0005 in order to ensure a smooth decay of the training loss. Training was for 100 epochs, with a batch size of 16, performed on 1-2 GPUs depending on availability. Due to training requiring a number of hours to complete, we did not repeat training runs in order to report a range in our chosen metric. 
we report the mean average precision at an intersection-over-union (IOU) threshold of 50\%, known as the mAP@50, in Table \ref{tab:RGZsource_detection}. This metric 
considers a prediction correct if the overlap between the predicted and actual object bounding boxes is at least 50\%, and is commonly used to evaluate object detection tasks.


Results showed that MSN and DINO are the best foundation models for this task. They outperformed training from scratch, even when the backbone was frozen, except for MSN on MGCLS. \newtext{Transformers pre-trained using multi-modal datasets (SigLIP and AM-RADIO) performed worse than the other ViTs, including MAE, on RGZ, and slightly better on MGCLS, although not as good as DINO and MSN. This is an interesting result, as both SigLIP and AM-RADIO were the best at classification. Representations learned through text-image pairs appear to be less relevant to object detection in images than they are to classification}.

Although the ResNet models did not outperform training from scratch when used as a frozen backbone, they equaled or exceeded that performance when fine-tuned. Detection on galaxies in the RGZ dataset appears to be a task particularly suited to ResNets, while Vision Transformers do very well at detecting compact sources in MGCLS. Generally, detection was very good on MGCLS for networks larger than ResNet-18. Possible reasons for the higher mAP@50 include the single category (compact source) and the tens of sources present in the images compared to the 1-6 sources in RGZ. Additionally, bounding boxes in RGZ can be quite large, including a lot of noise and making them harder to predict accurately.

These results show that transfer learning, starting with the pre-trained weights of a foundation model, is a good practice, as the right choice of model leads to better results even using a frozen backbone than training from scratch, and fine-tuning brings further improvements. It is important to ensure that properties of the architecture, such as patch size in the case of ViTs, are complementary to the data being used for the downstream task. 

\section{Best Practices Guidelines}\label{sec:bestpractices}

We do not claim that the the two datasets and two tasks examined here are completely representative of the many applications of deep learning for astrophysics. The lessons learned from considering various methods in using foundation models in these circumstances may be useful to practitioners, so we gather the results here into a general guideline.

As always, the dataset type -- characteristics of the images and type and number of available labels -- has the most impact on DST performance. These properties will inform the choice of foundation model, which in turn will determine how the dataset should be customized to complement the model itself. While GMNIST images are complementary enough to ImageNet and similar pre-trained backbones, a dataset like RGZ might benefit from a model with different pre-training. Models pre-trained on images with strong background patterns and textures, such as satellite or thermal imagery, or perhaps images from medical scans or microscopy could be good avenues to investigate. The labeling scheme of the dataset is also something to consider; the RGZ labels, while sensible from a scientific point of view, is not well suited to machine learning, as information about the number of radio components is obscured during scaling and conversion of the original images to PNG format.   

Besides selection of the foundation model for the backbone, there is also the choice of projector head for the downstream task. Clearly, some projector heads will have to be specifically designed for the task, such as Faster-RCNN for object detection. Other possibilities for detection and segmentation include YOLO, which is one of the best performing architectures on natural images, and can be used with a ResNet backbone. From a practical perspective, the choice of projector might also limit the choice of backbone; YOLO's latest versions can be adapted for ResNet backbones but not yet for ViTs. For simpler tasks like classification, the projector head is less important; we found that modifying it with an extra hidden layer or whitening layer only brought improvements of at most 5\%.

Fine-tuning the model, once selected, yields the most significant improvements. Model and dataset size determine the necessary time and resources; with many astrophysics datasets having less than 10K images, fine-tuning a model of a few million parameters should be feasible. An alternative to full fine-tuning is to only un-freeze the upper layers or blocks of a network. If computational resources are very limited, LoRA is an alternative that will generally be a few percent less accurate than full fine-tuning.

If fine-tuning does not reach the required performance metric, or if one wishes to generally improve the system as much as possible, one should carefully re-consider the data pre-processing. Together with custom dataset assembly and labeling, this is the most labor intensive step and requires the most specialized domain knowledge -- however it will also have the biggest impact on performance.  

For example, the simple task of resizing images to match the input dimensions of the model is often necessary, but can be done optimally. We saw that resizing images  can boost performance by up to 8\%, but that bigger was not necessarily better and was in fact worse when using ResNet-50. Furthermore, our example of RGZ classification showed that cropping images to focus on regions of interest was the next-best thing to fine-tuning. Considering the earlier discussion of the RGZ labeling scheme in Section \ref{sec:class_perform}, we also speculate that using different scalings of the original FITS images to include information about not only bright peaks but also flux islands could significantly improve distinction between visually similar classes like 2C-2P and 2C-3P (see Figure \ref{fig:sample_data} for example). One technique seen in the literature (\citealp{aniyan_classifying_2017}, used by \citet{maslej-kresnakova_morphological_2021} for example) is sigma-clipping, where noise is removed by setting all pixels below a certain threshold to zero. 

Note that in this work we do not discuss data augmentation, which is normally done during training to help the network learn more robust representations. Several works address augmentation schemes for both optical and radio images; \citet{hayat_self-supervised_2021} appendix D for optical, and \citet{slijepcevic_radio_2024} section 3.9 or \citet{maslej-kresnakova_morphological_2021} section 4.2 for radio. 

These general principles can help guide effective application of vision foundation models to astrophysical image data. Experimentation is unlikely to be a linear process; practitioners may need to iteratively revisit choices of foundation models, dataset preparation, and task-specific projector heads in order to achieve the desired result. 






\section{Conclusions}\label{sec:conclusion}

Our results demonstrate that state-of-the-art vision foundation models can immediately perform classification of optical galaxies and source detection of radio galaxies with F1 scores of 0.72 - 0.88. In the case of optical galaxy classification, all foundation models out-performed supervised training. F1 scores close to 0.87 with only 800 labeled training images (10\% of the total dataset) show that foundation models can be used to great effect for early-stage machine learning applications in scientific contexts. Of the models tested, SigLIP and AM-RADIO, pre-trained on multi-modal datasets, delivered the best performance, with Vision Transformers such as DINOv2 and MSN also excelling over ResNet-50, which is used or compared with in much of the astrophysics literature. Only MAE, which is optimized for reconstruction loss, failed to encode many task-relevant features for galaxy classification.  



Dataset characteristics strongly influenced model performance. For GMNIST, the alignment between the pre-training distribution and the task dataset likely contributed to the strong classification results, as galaxies occupied a significant portion of the image with minimal interference from noise or unrelated objects. Conversely, RGZ images, with small central sources and images dominated by noise and patterns from reconstruction, presented a clear distribution shift that limited the applicability of foundation models. In this case, supervised training outperformed all foundation models except SigLIP and AM-RADIO. While models retained some relevant features, such as bright peaks, we speculated that information about the flux islands, also relevant to classification, was lost due to the original pre-processing. Adapting dataset-specific pre-processing or label schemes 
could improve results, though significant domain expertise remains essential for working with radio astronomy datasets. 


 Source detection results further underscored the importance of aligning data to model architectures and tasks. When radio galaxy images were resized to ensure that transformer patch sizes were slightly smaller than average source dimensions, performance improved markedly. We saw that ViTs did well to detect compact sources in MGCLS, while ResNet was better at finding RGZ galaxies. Models pre-trained on multi-modal data usually did not perform as well as those pre-trained on only images, especially when freezing the backbone. Additionally, fine-tuning improvements were inconsistent and (in the case of MSN on RGZ) much smaller than might be expected. This situation might improve given a more optimal detection head architecture, or additional hyperparameter tuning.



Despite promising results, there remains a significant gap between the performance of foundation models on scientific datasets and their state-of-the-art results on natural image benchmarks. For example, ImageNet-1k classification achieves a top-1 accuracy of 92\%, and COCO object detection achieves an average mAP@50 of 0.73 across 90 classes. Closing this performance gap will require techniques that maximize results from limited labeled data and computational resources, a challenge particularly relevant in domains like astrophysics. Semi-supervised learning frameworks, such as those proposed by \citet{voloshynovskiy_variational_2020}, offer exciting possibilities to make use of the abundance of unlabeled data to enhance fine-tuning methods.  

In this study, we employed simple fine-tuning strategies, which, while effective, leave room for improvement. Future work will investigate advanced fine-tuning methods that integrate labeled and unlabeled data, potentially enabling more robust and resource-efficient application of vision foundation models to astrophysical image analysis. This iterative and adaptive approach will be essential for bridging the gap between cutting-edge machine learning methods and the unique demands of astronomical research.  

\section*{Data availability}\label{sec:data_availability}
GalaxyMNIST (\url{https://github.com/mwalmsley/galaxy_mnist}) and MGCLS (\url{https://doi.org/10.48479/7epd-w356}) are public datasets. Radio Galaxy Zoo (\url{https://radio.galaxyzoo.org/}) will soon have its first data release, and the dataset used here is available upon reasonable request. The code and instructions to reproduce these experiments is available at \url{https://github.com/elastufka/fm4astro}.

\section*{Acknowledgements}\label{sec:acknowledgements}
 
This work has been done in partnership of the
Swiss SKA consortium which is funded by the State Secretariat for Education, Research and Innovation (SERI).

MGCLS data products were provided by the South African Radio Astronomy Observatory and the MGCLS team and were derived from observations with the MeerKAT radio telescope. The MeerKAT telescope is operated by the South African Radio Astronomy Observatory, which is a facility of the National Research Foundation, an agency of the Department of Science and Innovation.

We would like to thank Philip Denzel for the RGZ object detection dataset.

OB and DP were supported by the SNF Sinergia grant CRSII5-193826 ``AstroSignals: A New Window on the Universe, with the New Generation of Large Radio-Astronomy Facilities''.

MD and VK were supported by the SNF Sinergia project (CRSII5-193716), ``Robust deep density models for high-energy particle physics and solar flare analysis (RODEM)''.

\bibliographystyle{aa}
\bibliography{FM_paper}

\begin{thebibliography}{76}
\expandafter\ifx\csname natexlab\endcsname\relax\def\natexlab#1{#1}\fi

\bibitem[{Aniyan \& Thorat(2017)}]{aniyan_classifying_2017}
Aniyan, A.~K. \& Thorat, K. 2017, The Astrophysical Journal Supplement Series,
  230, 20, publisher: The American Astronomical Society

\bibitem[{Assran {et~al.}(2022)Assran, Caron, Misra, Bojanowski, Bordes,
  Vincent, Joulin, Rabbat, \& Ballas}]{assran_masked_2022}
Assran, M., Caron, M., Misra, I., {et~al.} 2022, in Computer {Vision} –
  {ECCV} 2022: 17th {European} {Conference}, {Tel} {Aviv}, {Israel}, {October}
  23–27, 2022, {Proceedings}, {Part} {XXXI} (Berlin, Heidelberg:
  Springer-Verlag), 456--473

\bibitem[{Assran {et~al.}(2023)Assran, Duval, Misra, Bojanowski, Vincent,
  Rabbat, LeCun, \& Ballas}]{assran_self-supervised_2023}
Assran, M., Duval, Q., Misra, I., {et~al.} 2023, in 2023 {IEEE}/{CVF}
  {Conference} on {Computer} {Vision} and {Pattern} {Recognition} ({CVPR})
  (Vancouver, BC, Canada: IEEE), 15619--15629

\bibitem[{Balestriero \& LeCun(2024)}]{balestriero_how_2024}
Balestriero, R. \& LeCun, Y. 2024, in Forty-first {International} {Conference}
  on {Machine} {Learning}

\bibitem[{Bao {et~al.}(2022)Bao, Dong, Piao, \& Wei}]{bao_beit_2022}
Bao, H., Dong, L., Piao, S., \& Wei, F. 2022, in International {Conference} on
  {Learning} {Representations}

\bibitem[{Bardes {et~al.}(2022)Bardes, Ponce, \& LeCun}]{bardes_vicreg_2022}
Bardes, A., Ponce, J., \& LeCun, Y. 2022, in International {Conference} on
  {Learning} {Representations}

\bibitem[{Becker {et~al.}(2021)Becker, Vaccari, Prescott, \&
  Grobler}]{becker_cnn_2021}
Becker, B., Vaccari, M., Prescott, M., \& Grobler, T. 2021, Monthly Notices of
  the Royal Astronomical Society, 503, 1828, aDS Bibcode: 2021MNRAS.503.1828B

\bibitem[{Burke {et~al.}(2019)Burke, Aleo, Chen, Liu, Peterson, Sembroski, \&
  Lin}]{burke_deblending_2019}
Burke, C.~J., Aleo, P.~D., Chen, Y.-C., {et~al.} 2019, Monthly Notices of the
  Royal Astronomical Society, 490, 3952

\bibitem[{Cao {et~al.}(2024)Cao, Xu, Deng, Deng, Yang, Liu, \&
  Zhou}]{cao_galaxy_2024}
Cao, J., Xu, T., Deng, Y., {et~al.} 2024, Astronomy \& Astrophysics, 683, A42,
  publisher: EDP Sciences

\bibitem[{Caron {et~al.}(2020)Caron, Misra, Mairal, Goyal, Bojanowski, \&
  Joulin}]{caron_unsupervised_2020}
Caron, M., Misra, I., Mairal, J., {et~al.} 2020, in Advances in {Neural}
  {Information} {Processing} {Systems}, Vol.~33 (Curran Associates, Inc.),
  9912--9924

\bibitem[{Caron {et~al.}(2021)Caron, Touvron, Misra, Jegou, Mairal, Bojanowski,
  \& Joulin}]{caron_emerging_2021}
Caron, M., Touvron, H., Misra, I., {et~al.} 2021, in 2021 {IEEE}/{CVF}
  {International} {Conference} on {Computer} {Vision} ({ICCV}), 9630--9640,
  iSSN: 2380-7504

\bibitem[{Chen {et~al.}(2024{\natexlab{a}})Chen, Bianco, Tolley, Spinelli,
  Forero-Sanchez, \& Kneib}]{chen_stability_2024}
Chen, T., Bianco, M., Tolley, E., {et~al.} 2024{\natexlab{a}}, Monthly Notices
  of the Royal Astronomical Society, 532, 2615

\bibitem[{Chen {et~al.}(2020)Chen, Kornblith, Norouzi, \&
  Hinton}]{chen_simple_2020}
Chen, T., Kornblith, S., Norouzi, M., \& Hinton, G. 2020, in International
  conference on machine learning (PMLR), 1597--1607

\bibitem[{Chen {et~al.}(2024{\natexlab{b}})Chen, Ding, Wang, Xin, Mo, Wang,
  Han, Luo, Zeng, \& Wang}]{chen_context_2024}
Chen, X., Ding, M., Wang, X., {et~al.} 2024{\natexlab{b}}, International
  Journal of Computer Vision, 132, 208

\bibitem[{Cohen \& Lu(2021)}]{cohen_diffusion-based_2021}
Cohen, M. \& Lu, W. 2021, Astronomy and Computing, 37, 100507

\bibitem[{Cornu {et~al.}(2024)Cornu, Salomé, Semelin, Marchal, Freundlich,
  Aicardi, Lu, Sainton, Mertens, Combes, \& Tasse}]{cornu_yolo-cianna_2024}
Cornu, D., Salomé, P., Semelin, B., {et~al.} 2024, {YOLO}-{CIANNA}: {Galaxy}
  detection with deep learning in radio data. {I}. {A} new {YOLO}-inspired
  source detection method applied to the {SKAO} {SDC1}, arXiv:2402.05925

\bibitem[{Dagli(2023)}]{dagli_astroformer_2023}
Dagli, R. 2023, Astroformer: {More} {Data} {Might} not be all you need for
  {Classification}, arXiv:2304.05350 [cs]

\bibitem[{Deng {et~al.}(2009)Deng, Dong, Socher, Li, Li, \&
  Fei-Fei}]{deng_imagenet_2009}
Deng, J., Dong, W., Socher, R., {et~al.} 2009, in 2009 {IEEE} {Conference} on
  {Computer} {Vision} and {Pattern} {Recognition}, 248--255, iSSN: 1063-6919

\bibitem[{Dieleman {et~al.}(2015)Dieleman, Willett, \&
  Dambre}]{dieleman_rotation-invariant_2015}
Dieleman, S., Willett, K.~W., \& Dambre, J. 2015, Monthly Notices of the Royal
  Astronomical Society, 450, 1441

\bibitem[{Dosovitskiy {et~al.}(2021)Dosovitskiy, Beyer, Kolesnikov,
  Weissenborn, Zhai, Unterthiner, Dehghani, Minderer, Heigold, Gelly,
  Uszkoreit, \& Houlsby}]{dosovitskiy_image_2021}
Dosovitskiy, A., Beyer, L., Kolesnikov, A., {et~al.} 2021, An {Image} is
  {Worth} 16x16 {Words}: {Transformers} for {Image} {Recognition} at {Scale},
  arXiv:2010.11929

\bibitem[{Drozdova {et~al.}(2023)Drozdova, Kinakh, Bait, Taran, Lastufka,
  Dessauges-Zavadsky, Holotyak, Schaerer, \&
  Voloshynovskiy}]{drozdova_radio-astronomical_2023}
Drozdova, M., Kinakh, V., Bait, O., {et~al.} 2023, Astronomy \& Astrophysics,
  683

\bibitem[{Farias {et~al.}(2020)Farias, Ortiz, Damke, Jaque~Arancibia, \&
  Solar}]{farias_mask_2020}
Farias, H., Ortiz, D., Damke, G., Jaque~Arancibia, M., \& Solar, M. 2020,
  Astronomy and Computing, 33, 100420

\bibitem[{Grill {et~al.}(2020)Grill, Strub, Altché, Tallec, Richemond,
  Buchatskaya, Doersch, Pires, Guo, Azar, Piot, Kavukcuoglu, Munos, \&
  Valko}]{grill_bootstrap_2020}
Grill, J.-B., Strub, F., Altché, F., {et~al.} 2020, in Proceedings of the 34th
  {International} {Conference} on {Neural} {Information} {Processing}
  {Systems}, {NIPS} '20 (Red Hook, NY, USA: Curran Associates Inc.),
  21271--21284

\bibitem[{Ha \& Schmidhuber(2018)}]{ha_world_2018}
Ha, D. \& Schmidhuber, J. 2018, World {Models}, arXiv:1803.10122 [cs]

\bibitem[{Hausen \& Robertson(2022)}]{hausen_partial-attribution_2022}
Hausen, R. \& Robertson, B. 2022, Partial-{Attribution} {Instance}
  {Segmentation} for {Astronomical} {Source} {Detection} and {Deblending},
  arXiv:2201.04714 [astro-ph]

\bibitem[{Hayat {et~al.}(2021)Hayat, Stein, Harrington, Lukić, \&
  Mustafa}]{hayat_self-supervised_2021}
Hayat, M.~A., Stein, G., Harrington, P., Lukić, Z., \& Mustafa, M. 2021, The
  Astrophysical Journal Letters, 911, L33, publisher: The American Astronomical
  Society

\bibitem[{He {et~al.}(2022)He, Chen, Xie, Li, Dollár, \&
  Girshick}]{he_masked_2022}
He, K., Chen, X., Xie, S., {et~al.} 2022, in 2022 {IEEE}/{CVF} {Conference} on
  {Computer} {Vision} and {Pattern} {Recognition} ({CVPR}), 15979--15988, iSSN:
  2575-7075

\bibitem[{He {et~al.}(2016)He, Zhang, Ren, \& Sun}]{he_deep_2016}
He, K., Zhang, X., Ren, S., \& Sun, J. 2016, in 2016 {IEEE} {Conference} on
  {Computer} {Vision} and {Pattern} {Recognition} ({CVPR}), 770--778, iSSN:
  1063-6919

\bibitem[{Hui {et~al.}(2022)Hui, Robert~Jia, Li, \& Wang}]{hui_galaxy_2022}
Hui, W., Robert~Jia, Z., Li, H., \& Wang, Z. 2022, Journal of Physics:
  Conference Series, 2402, 012009

\bibitem[{Jia {et~al.}(2021)Jia, Yang, Xia, Chen, Parekh, Pham, Le, Sung, Li,
  \& Duerig}]{jia_scaling_2021}
Jia, C., Yang, Y., Xia, Y., {et~al.} 2021, in International conference on
  machine learning (PMLR), 4904--4916

\bibitem[{Jia {et~al.}(2023)Jia, Zheng, Wang, \& Yang}]{jia_deep_2023}
Jia, P., Zheng, Y., Wang, M., \& Yang, Z. 2023, Astronomy and Computing, 42,
  100687

\bibitem[{Kalapos \& Gyires-Tóth(2024)}]{kalapos_whitening_2024}
Kalapos, A. \& Gyires-Tóth, B. 2024, Whitening {Consistently} {Improves}
  {Self}-{Supervised} {Learning}, arXiv:2408.07519 [cs]

\bibitem[{Kalvankar {et~al.}(2021)Kalvankar, Pandit, \&
  Parwate}]{kalvankar_galaxy_2021}
Kalvankar, S., Pandit, H., \& Parwate, P. 2021, Galaxy {Morphology}
  {Classification} using {EfficientNet} {Architectures}, arXiv:2008.13611 [cs]

\bibitem[{Kirillov {et~al.}(2023)Kirillov, Mintun, Ravi, Mao, Rolland,
  Gustafson, Xiao, Whitehead, Berg, Lo, \& {others}}]{kirillov_segment_2023}
Kirillov, A., Mintun, E., Ravi, N., {et~al.} 2023, in Proceedings of the
  {IEEE}/{CVF} {International} {Conference} on {Computer} {Vision}, 4015--4026

\bibitem[{Kumar {et~al.}(2023)Kumar, Sarker, \& Islam}]{kumar_vision_2023}
Kumar, R., Sarker, M.~K., \& Islam, S.~R. 2023, in Deep {Learning} {Theory} and
  {Applications}, ed. D.~Conte, A.~Fred, O.~Gusikhin, \& C.~Sansone (Cham:
  Springer Nature Switzerland), 115--125

\bibitem[{Lastufka {et~al.}(2024)Lastufka, Bait, Taran, Drozdova, Kinakh,
  Piras, Audard, Dessauges-Zavadsky, Holotyak, Schaerer, \&
  Voloshynovskiy}]{lastufka_self-supervised_2024}
Lastufka, E., Bait, O., Taran, O., {et~al.} 2024, Astronomy \& Astrophysics,
  publisher: EDP Sciences

\bibitem[{Li {et~al.}(2022{\natexlab{a}})Li, Zhang, Zhang, Yang, Li, Zhong,
  Wang, Yuan, Zhang, Hwang, Chang, \& Gao}]{li_grounded_2022}
Li, L.~H., Zhang, P., Zhang, H., {et~al.} 2022{\natexlab{a}}, in 2022
  {IEEE}/{CVF} {Conference} on {Computer} {Vision} and {Pattern} {Recognition}
  ({CVPR}) (New Orleans, LA, USA: IEEE), 10955--10965

\bibitem[{Li {et~al.}(2022{\natexlab{b}})Li, Mao, Girshick, \&
  He}]{li_exploring_2022}
Li, Y., Mao, H., Girshick, R., \& He, K. 2022{\natexlab{b}}, in European
  conference on computer vision (Springer), 280--296

\bibitem[{Li {et~al.}(2021)Li, Yu, Xiao, Long, \& Cui}]{li_detection_2021}
Li, Z., Yu, C., Xiao, J., Long, M., \& Cui, C. 2021, Astronomy and Computing,
  36, 100482

\bibitem[{Lin {et~al.}(2022)Lin, Liao, Huang, Kuo, \& Ou}]{lin_galaxy_2022}
Lin, J. Y.-Y., Liao, S.-M., Huang, H.-J., Kuo, W.-T., \& Ou, O. H.-M. 2022,
  Galaxy {Morphological} {Classification} with {Efficient} {Vision}
  {Transformer}, arXiv:2110.01024

\bibitem[{Lochner \& Bassett(2021)}]{lochner_astronomaly_2021}
Lochner, M. \& Bassett, B.~A. 2021, Astronomy and Computing, 36, 100481,
  arXiv:2010.11202 [astro-ph]

\bibitem[{Marcel \& Rodriguez(2010)}]{marcel_torchvision_2010}
Marcel, S. \& Rodriguez, Y. 2010, in Proceedings of the 18th {ACM}
  international conference on {Multimedia}, {MM} '10 (New York, NY, USA:
  Association for Computing Machinery), 1485--1488

\bibitem[{Maslej-Krešňáková {et~al.}(2021)Maslej-Krešňáková,
  El Bouchefry, \& Butka}]{maslej-kresnakova_morphological_2021}
Maslej-Krešňáková, V., El Bouchefry, K., \& Butka, P. 2021, Monthly
  Notices of the Royal Astronomical Society, 505, 1464

\bibitem[{Merz {et~al.}(2023)Merz, Liu, Burke, Aleo, Liu, Carrasco Kind,
  Kindratenko, \& Liu}]{merz_detection_2023}
Merz, G., Liu, Y., Burke, C.~J., {et~al.} 2023, Monthly Notices of the Royal
  Astronomical Society, 526, 1122

\bibitem[{Müller \& Hutter(2021)}]{muller_trivialaugment_2021}
Müller, S.~G. \& Hutter, F. 2021, in Proceedings of the {IEEE}/{CVF}
  international conference on computer vision, 774--782

\bibitem[{Ndung’u {et~al.}(2023)Ndung’u, Grobler, Wijnholds, Karastoyanova,
  \& Azzopardi}]{ndungu_advances_2023}
Ndung’u, S., Grobler, T., Wijnholds, S.~J., Karastoyanova, D., \& Azzopardi,
  G. 2023, New Astronomy Reviews, 97, 101685

\bibitem[{Oquab {et~al.}(2024)Oquab, Darcet, Moutakanni, Vo, Szafraniec,
  Khalidov, Fernandez, HAZIZA, Massa, El-Nouby, Assran, Ballas, Galuba, Howes,
  Huang, Li, Misra, Rabbat, Sharma, Synnaeve, Xu, Jegou, Mairal, Labatut,
  Joulin, \& Bojanowski}]{oquab_dinov2_2024}
Oquab, M., Darcet, T., Moutakanni, T., {et~al.} 2024, Transactions on Machine
  Learning Research

\bibitem[{Pandya {et~al.}(2023)Pandya, Patel, O, \& Blazek}]{pandya_e2_2023}
Pandya, S., Patel, P., O, F., \& Blazek, J. 2023, E(2) {Equivariant} {Neural}
  {Networks} for {Robust} {Galaxy} {Morphology} {Classification},
  arXiv:2311.01500

\bibitem[{Porter \& Scaife(2023)}]{porter_mirabest_2023}
Porter, F. A.~M. \& Scaife, A. M.~M. 2023, RAS Techniques and Instruments, 2,
  293

\bibitem[{Radford {et~al.}(2021)Radford, Kim, Hallacy, Ramesh, Goh, Agarwal,
  Sastry, Askell, Mishkin, Clark, \& {others}}]{radford_learning_2021}
Radford, A., Kim, J.~W., Hallacy, C., {et~al.} 2021, in International
  conference on machine learning (PMLR), 8748--8763

\bibitem[{Ranzinger {et~al.}(2024)Ranzinger, Heinrich, Kautz, \&
  Molchanov}]{ranzinger_am-radio_2024}
Ranzinger, M., Heinrich, G., Kautz, J., \& Molchanov, P. 2024, in Proceedings
  of the {IEEE}/{CVF} {Conference} on {Computer} {Vision} and {Pattern}
  {Recognition}, 12490--12500

\bibitem[{Redmon {et~al.}(2016)Redmon, Divvala, Girshick, \&
  Farhadi}]{redmon_you_2016}
Redmon, J., Divvala, S., Girshick, R., \& Farhadi, A. 2016, in 2016 {IEEE}
  {Conference} on {Computer} {Vision} and {Pattern} {Recognition} ({CVPR}) (Las
  Vegas, NV, USA: IEEE), 779--788

\bibitem[{Reiman \& Göhre(2019)}]{reiman_deblending_2019}
Reiman, D.~M. \& Göhre, B.~E. 2019, Monthly Notices of the Royal Astronomical
  Society, 485, 2617

\bibitem[{Ren {et~al.}(2017)Ren, He, Girshick, \& Sun}]{ren_faster_2017}
Ren, S., He, K., Girshick, R., \& Sun, J. 2017, IEEE Transactions on Pattern
  Analysis and Machine Intelligence, 39, 1137

\bibitem[{Riggi {et~al.}(2023)Riggi, Magro, Sortino, De~Marco, Bordiu,
  Cecconello, Hopkins, Marvil, Umana, Sciacca, Vitello, Bufano, Ingallinera,
  Fiameni, Spampinato, \& Zarb~Adami}]{riggi_astronomical_2023}
Riggi, S., Magro, D., Sortino, R., {et~al.} 2023, Astronomy and Computing, 42,
  100682

\bibitem[{Schmidt {et~al.}(2022)Schmidt, Geyer, Fröse, Blomenkamp, Brüggen,
  Gasperin, Elsässer, \& Rhode}]{schmidt_deep_2022}
Schmidt, K., Geyer, F., Fröse, S., {et~al.} 2022, Astronomy \& Astrophysics,
  664, A134, publisher: EDP Sciences

\bibitem[{Slijepcevic {et~al.}(2024)Slijepcevic, Scaife, Walmsley, Bowles,
  Wong, Shabala, \& White}]{slijepcevic_radio_2024}
Slijepcevic, I.~V., Scaife, A. M.~M., Walmsley, M., {et~al.} 2024, RAS
  Techniques and Instruments, 3, 19

\bibitem[{Sortino {et~al.}(2023)Sortino, Magro, Fiameni, Sciacca, Riggi,
  DeMarco, Spampinato, Hopkins, Bufano, Schillirò, Bordiu, \&
  Pino}]{sortino_radio_2023}
Sortino, R., Magro, D., Fiameni, G., {et~al.} 2023, Experimental Astronomy,
  arXiv:2303.04506 [cs]

\bibitem[{Taran {et~al.}(2023)Taran, Bait, Dessauges-Zavadsky, Holotyak,
  Schaerer, \& Voloshynovskiy}]{taran_challenging_2023}
Taran, O., Bait, O., Dessauges-Zavadsky, M., {et~al.} 2023, Astronomy \&
  Astrophysics, 674, A161, publisher: EDP Sciences

\bibitem[{Tishby {et~al.}(2000)Tishby, Pereira, \&
  Bialek}]{tishby_information_2000}
Tishby, N., Pereira, F.~C., \& Bialek, W. 2000, The information bottleneck
  method, arXiv:physics/0004057

\bibitem[{Urechiatu \& Frincu(2024)}]{urechiatu_improved_2024}
Urechiatu, R. \& Frincu, M. 2024, Universe, 10, 230, number: 6 Publisher:
  Multidisciplinary Digital Publishing Institute

\bibitem[{Vafaei Sadr {et~al.}(2019)Vafaei Sadr, Vos, Bassett, Hosenie,
  Oozeer, \& Lochner}]{vafaeisadr_deepsource_2019}
Vafaei Sadr, A., Vos, E.~E., Bassett, B.~A., {et~al.} 2019, Monthly Notices of
  the Royal Astronomical Society, 484, 2793

\bibitem[{Villar {et~al.}(2021)Villar, Cranmer, Berger, Contardo, Ho,
  Hosseinzadeh, \& Lin}]{villar_deep-learning_2021}
Villar, V.~A., Cranmer, M., Berger, E., {et~al.} 2021, The Astrophysical
  Journal Supplement Series, 255, 24, publisher: The American Astronomical
  Society

\bibitem[{Voloshynovskiy {et~al.}(2020)Voloshynovskiy, Taran, Kondah, Holotyak,
  \& Rezende}]{voloshynovskiy_variational_2020}
Voloshynovskiy, S., Taran, O., Kondah, M., Holotyak, T., \& Rezende, D. 2020,
  Entropy, 22, 943, number: 9 Publisher: Multidisciplinary Digital Publishing
  Institute

\bibitem[{Vos {et~al.}(2019)Vos, Francois~Luus, Finlay, \&
  Bassett}]{vos_generative_2019}
Vos, E.~E., Francois~Luus, P.~S., Finlay, C.~J., \& Bassett, B.~A. 2019, in
  2019 {IEEE} 29th {International} {Workshop} on {Machine} {Learning} for
  {Signal} {Processing} ({MLSP}), 1--6, iSSN: 1551-2541

\bibitem[{Walmsley {et~al.}(2022)Walmsley, Lintott, Géron, Kruk, Krawczyk,
  Willett, Bamford, Kelvin, Fortson, Gal, Keel, Masters, Mehta, Simmons,
  Smethurst, Smith, Baeten, \& Macmillan}]{walmsley_galaxy_2022}
Walmsley, M., Lintott, C., Géron, T., {et~al.} 2022, Monthly Notices of the
  Royal Astronomical Society, 509, 3966

\bibitem[{Wei {et~al.}(2024)Wei, Lu, Dai, Liang, Hao, Zhang, \&
  Zhang}]{wei_galaxy_2024}
Wei, S., Lu, W., Dai, W., {et~al.} 2024, The Astronomical Journal, 167, 29

\bibitem[{Wong {et~al.}(2024)Wong, Garon, Alger, Rudnick, Shabala, Willett,
  Banfield, Andernach, Norris, Swan, \& {others}}]{wong_radio_2024}
Wong, O.~I., Garon, A., Alger, M., {et~al.} 2024, Monthly Notices of the Royal
  Astronomical Society, stae2790, publisher: Oxford University Press

\bibitem[{Wu {et~al.}(2019)Wu, Wong, Rudnick, Shabala, Alger, Banfield, Ong,
  White, Garon, Norris, Andernach, Tate, Lukic, Tang, Schawinski, \&
  Diakogiannis}]{wu_radio_2019}
Wu, C., Wong, O.~I., Rudnick, L., {et~al.} 2019, Monthly Notices of the Royal
  Astronomical Society, 482, 1211

\bibitem[{Yu {et~al.}(2022)Yu, Wang, Vasudevan, Yeung, Seyedhosseini, \&
  Wu}]{yu_coca_2022}
Yu, J., Wang, Z., Vasudevan, V., {et~al.} 2022, Transactions on Machine
  Learning Research

\bibitem[{Zbontar {et~al.}(2021)Zbontar, Jing, Misra, LeCun, \&
  Deny}]{zbontar_barlow_2021-1}
Zbontar, J., Jing, L., Misra, I., LeCun, Y., \& Deny, S. 2021, in Proceedings
  of the 38th {International} {Conference} on {Machine} {Learning} (PMLR),
  12310--12320, iSSN: 2640-3498

\bibitem[{Zhai {et~al.}(2022)Zhai, Kolesnikov, Houlsby, \&
  Beyer}]{zhai_scaling_2022}
Zhai, X., Kolesnikov, A., Houlsby, N., \& Beyer, L. 2022, in Proceedings of the
  {IEEE}/{CVF} conference on computer vision and pattern recognition,
  12104--12113

\bibitem[{Zhai {et~al.}(2023)Zhai, Mustafa, Kolesnikov, \&
  Beyer}]{zhai_sigmoid_2023}
Zhai, X., Mustafa, B., Kolesnikov, A., \& Beyer, L. 2023, in Proceedings of the
  {IEEE}/{CVF} {International} {Conference} on {Computer} {Vision},
  11975--11986

\bibitem[{Zhou {et~al.}(2022)Zhou, Wei, Wang, Shen, Xie, Yuille, \&
  Kong}]{zhou_image_2022}
Zhou, J., Wei, C., Wang, H., {et~al.} 2022, in International {Conference} on
  {Learning} {Representations}

\bibitem[{Zhou {et~al.}(2023)Zhou, Gong, Deng, Zhang, Yue, \&
  Chen}]{zhou_foreground_2023}
Zhou, X., Gong, Y., Deng, F., {et~al.} 2023, Monthly Notices of the Royal
  Astronomical Society, 521, 278

\bibitem[{Zhu {et~al.}(2019)Zhu, Dai, Bian, Chen, Chen, \&
  Hu}]{zhu_galaxy_2019}
Zhu, X.-P., Dai, J.-M., Bian, C.-J., {et~al.} 2019, Astrophysics and Space
  Science, 364, 55

\end{thebibliography}

\begin{appendix} 
\onecolumn






\section{Attention Visualization}\label{sec:attention}

\begin{figure}[ht]
\centering
\subfloat{\includegraphics[width=\linewidth,trim={3.6cm 5.5cm 3.3cm 4.75cm},clip]{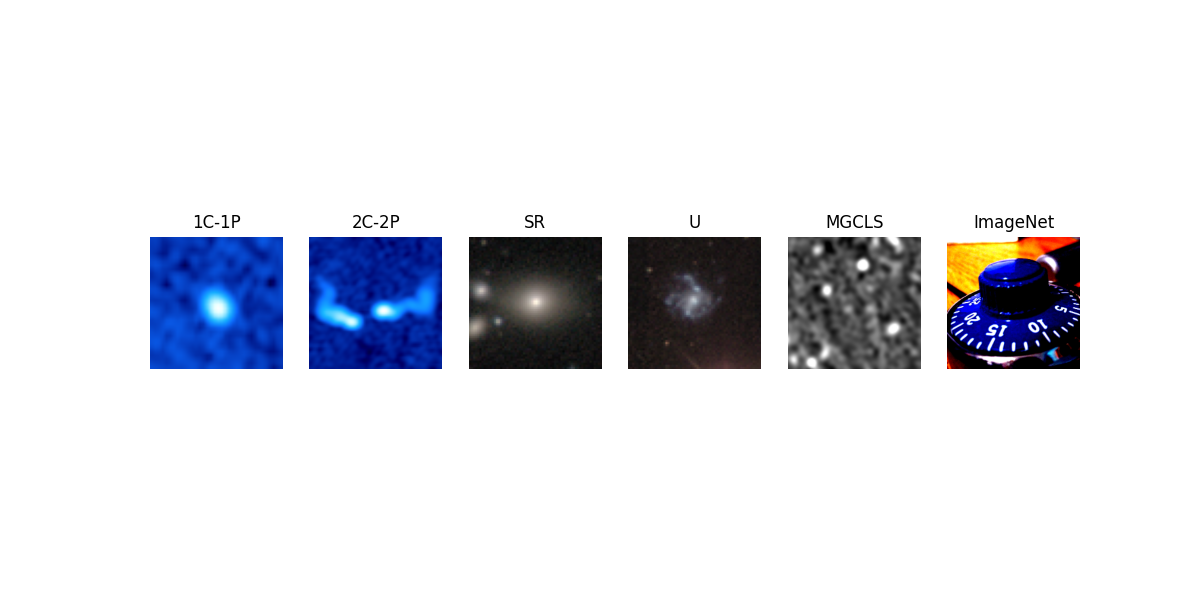}}
\caption{Images for the following attention maps} \label{fig:attentionims}
\end{figure}

\begin{figure}[!h]
\centering
\subfloat{\includegraphics[width=0.49\linewidth,trim={1.5cm 2cm 2cm 1.8cm},clip]{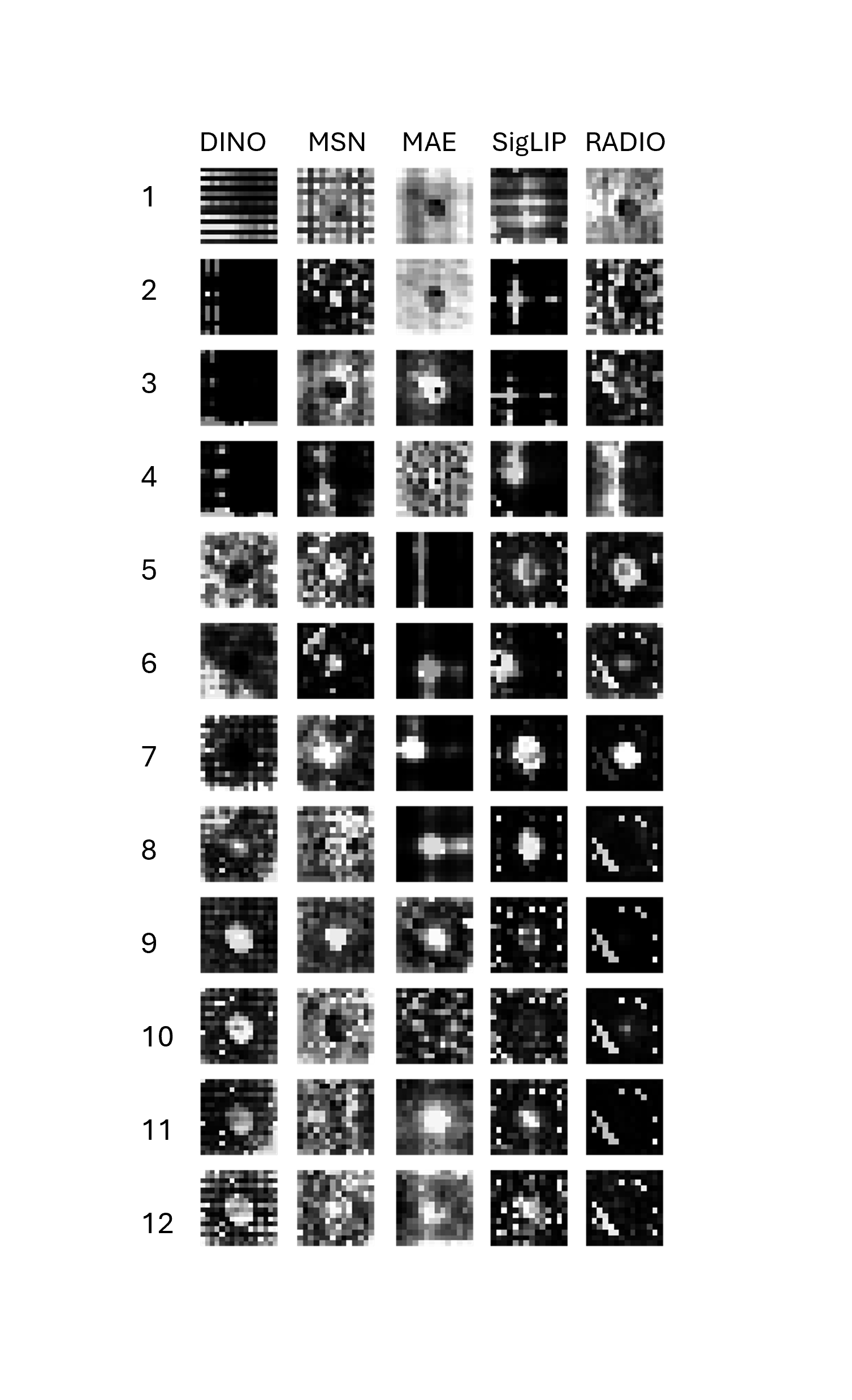}}
\subfloat{\includegraphics[width=0.49\linewidth,trim={1.5cm 2cm 2cm 1.8cm},clip]{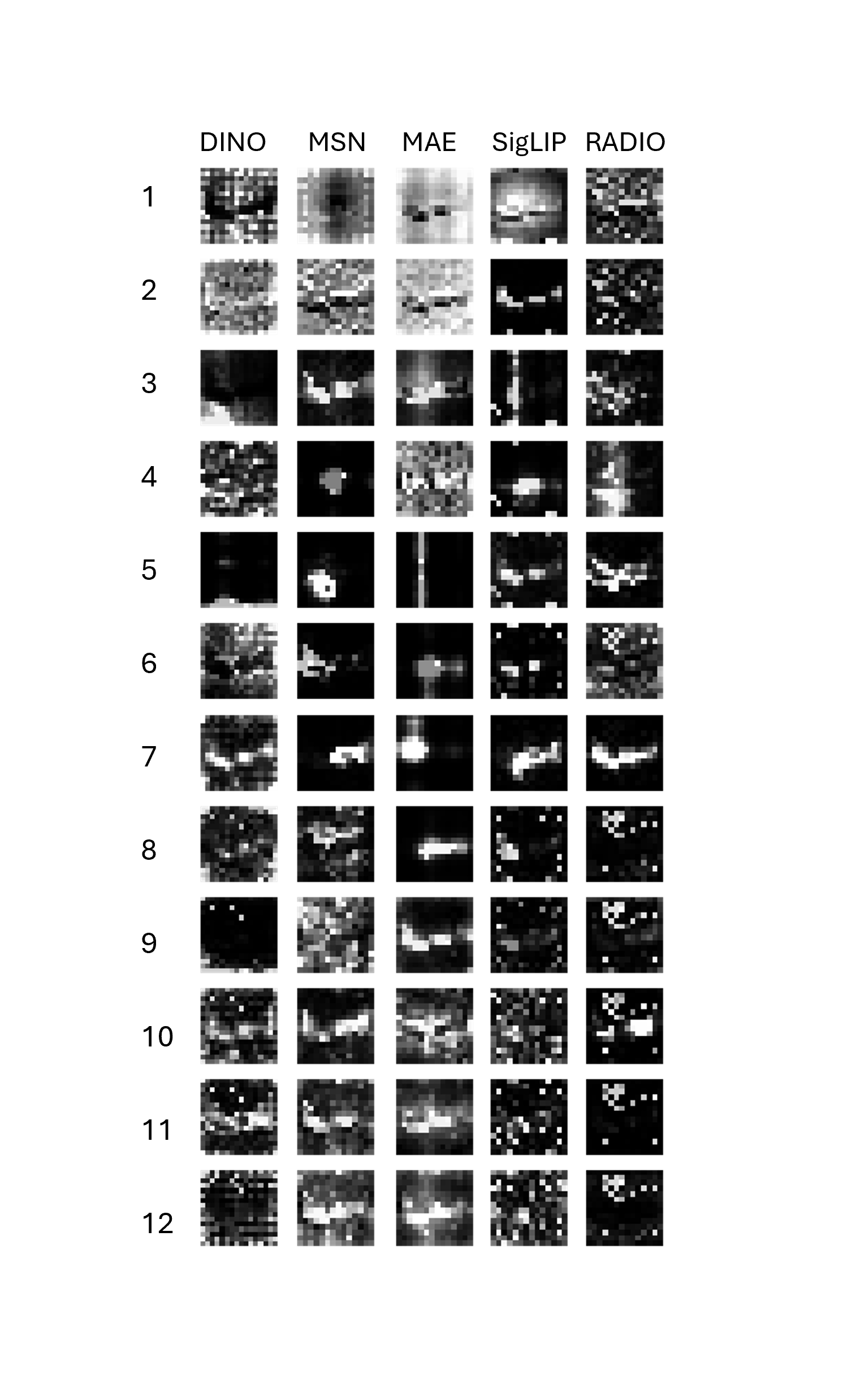}}
\caption{Attention maps for all models, RGZ compact source (left, labeled in the dataset as 1C-1P) and extended source (right, labeled in the dataset as 2C-2P), shown for each model and each attention block.} \label{fig:rgzattention}
\end{figure}

\begin{figure*}[h]
\centering
\subfloat{\includegraphics[width=0.49\linewidth,trim={1.5cm 2cm 2cm 1.8cm},clip]{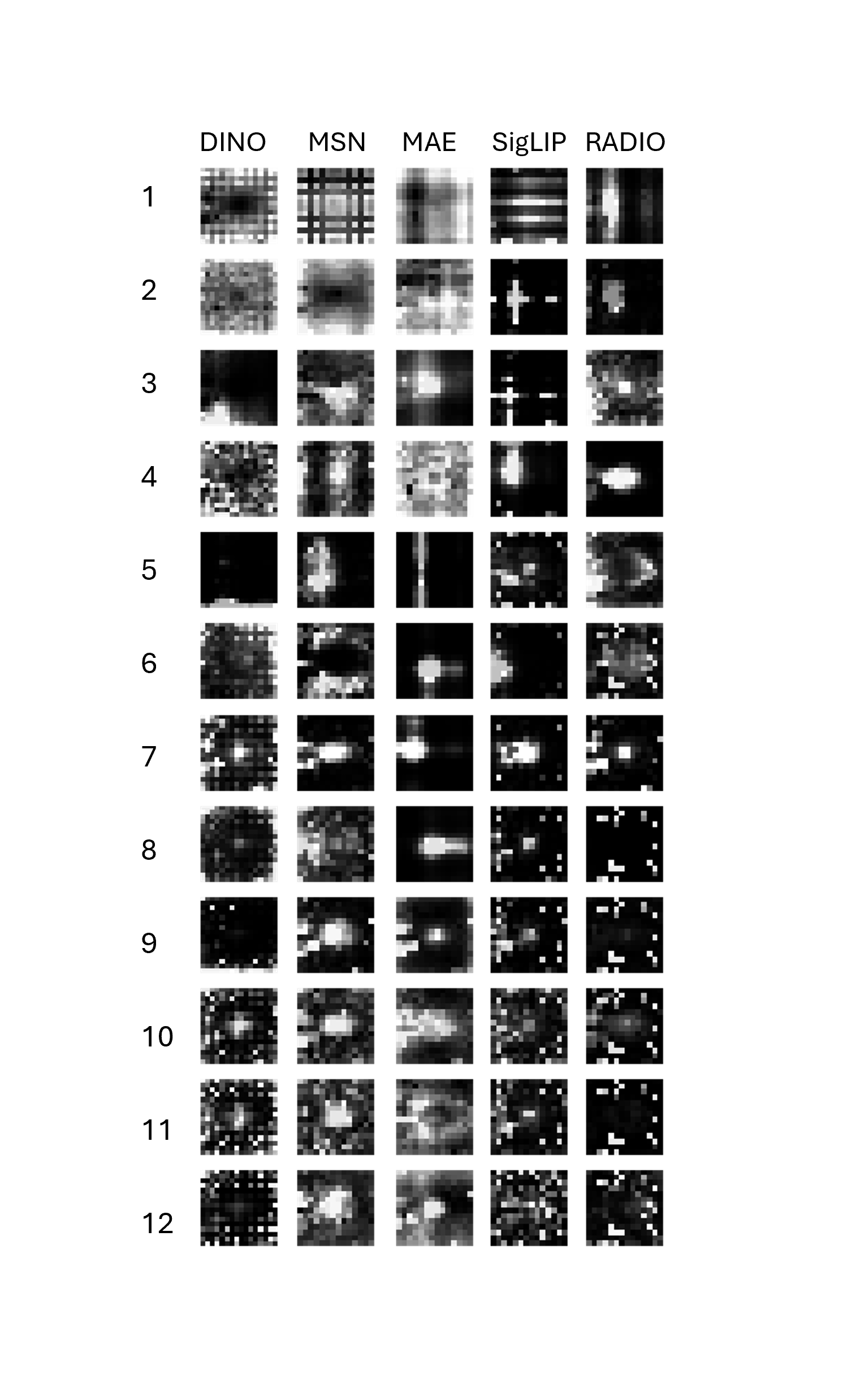}}
\subfloat{\includegraphics[width=0.49\linewidth,trim={1.5cm 2cm 2cm 1.8cm},clip]{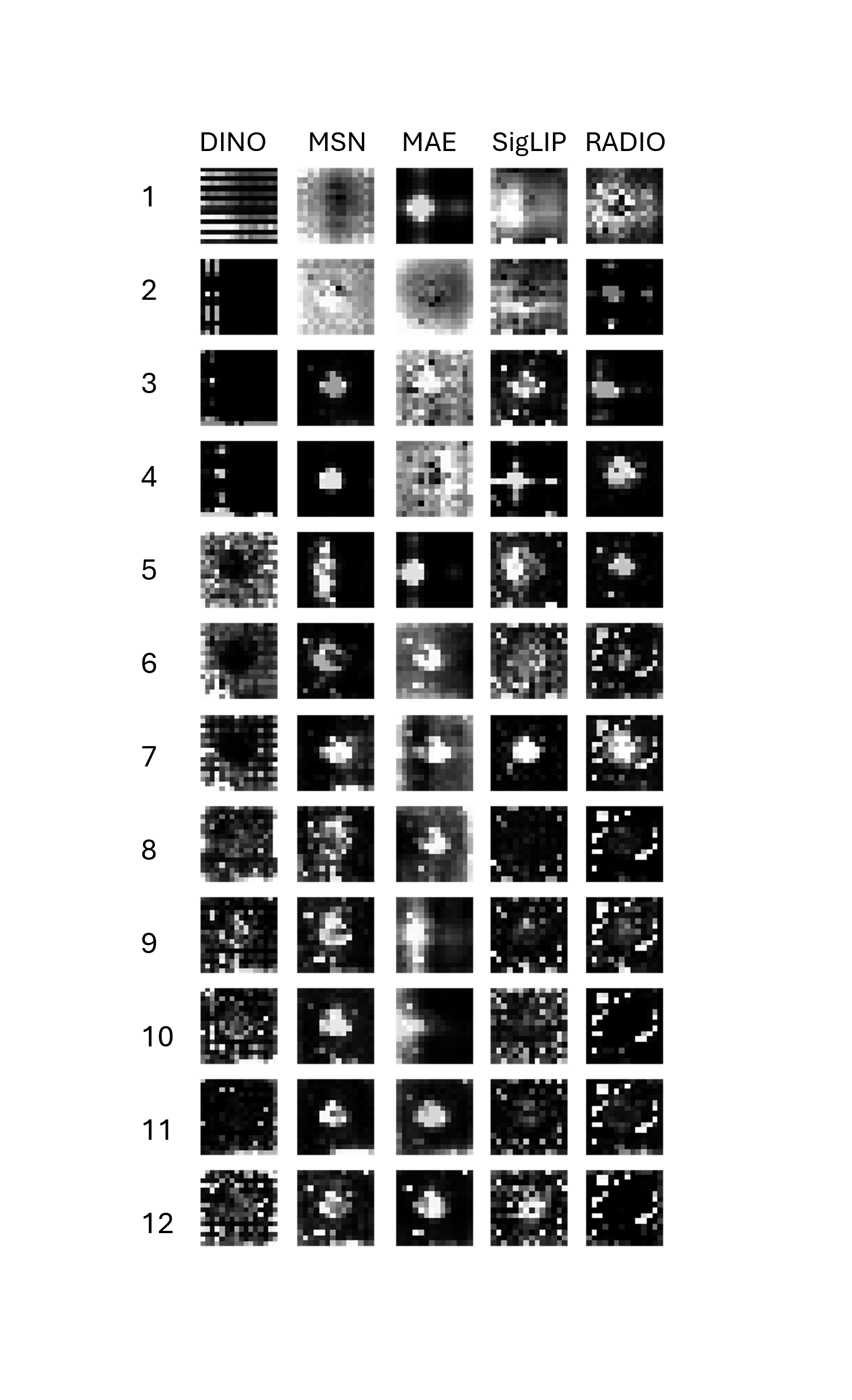}}
\caption{Attention maps for all models, GMNIST smooth-and-round galaxy (left) and unbarred spiral (right), shown for each model and each attention block.} \label{fig:gmnistattention}
\end{figure*}

\begin{figure*}[h]
\centering
\subfloat{\includegraphics[width=0.49\linewidth,trim={1.5cm 2cm 2cm 1.8cm},clip]{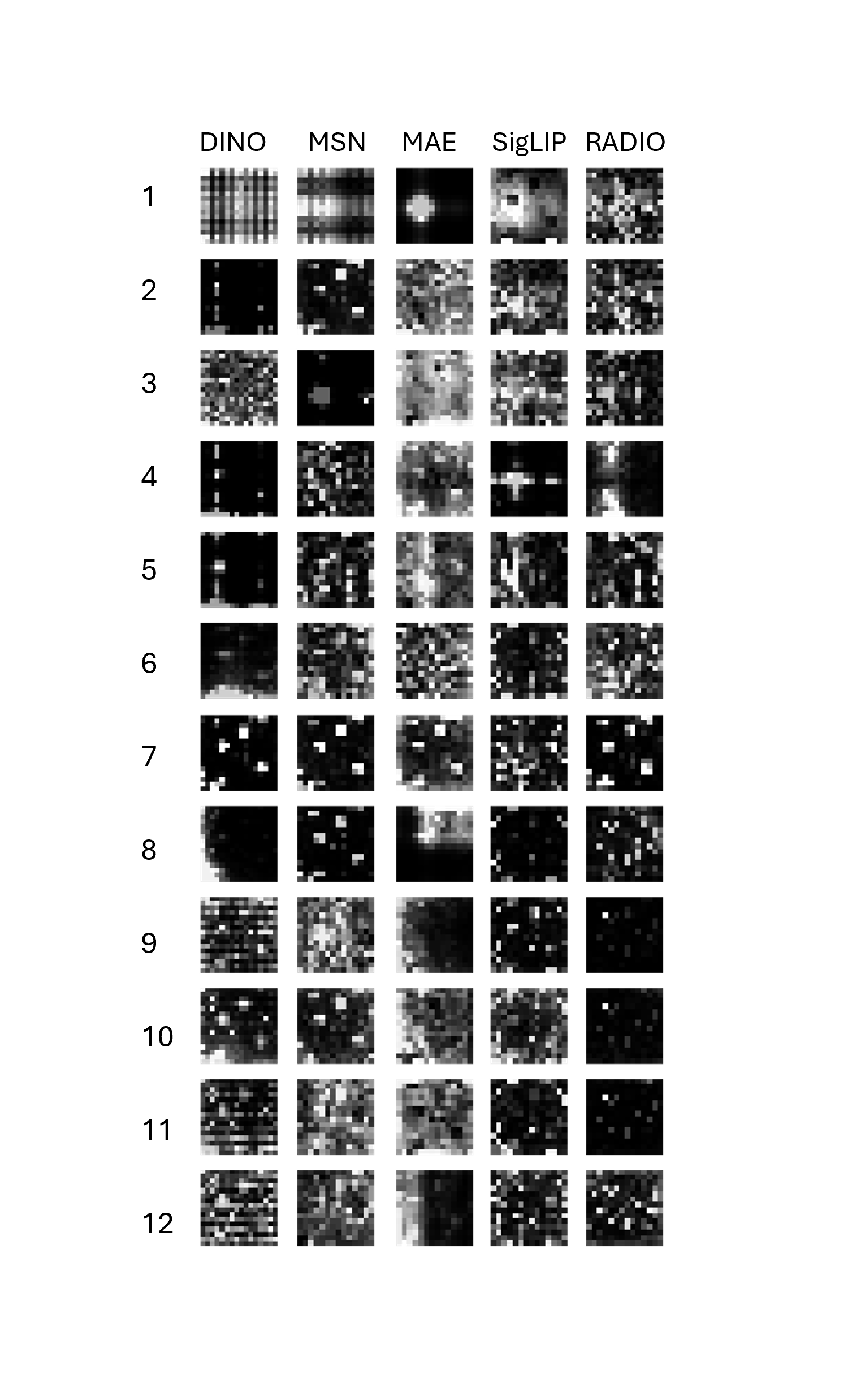}}
\subfloat{\includegraphics[width=0.49\linewidth,trim={1.5cm 2cm 2cm 1.8cm},clip]{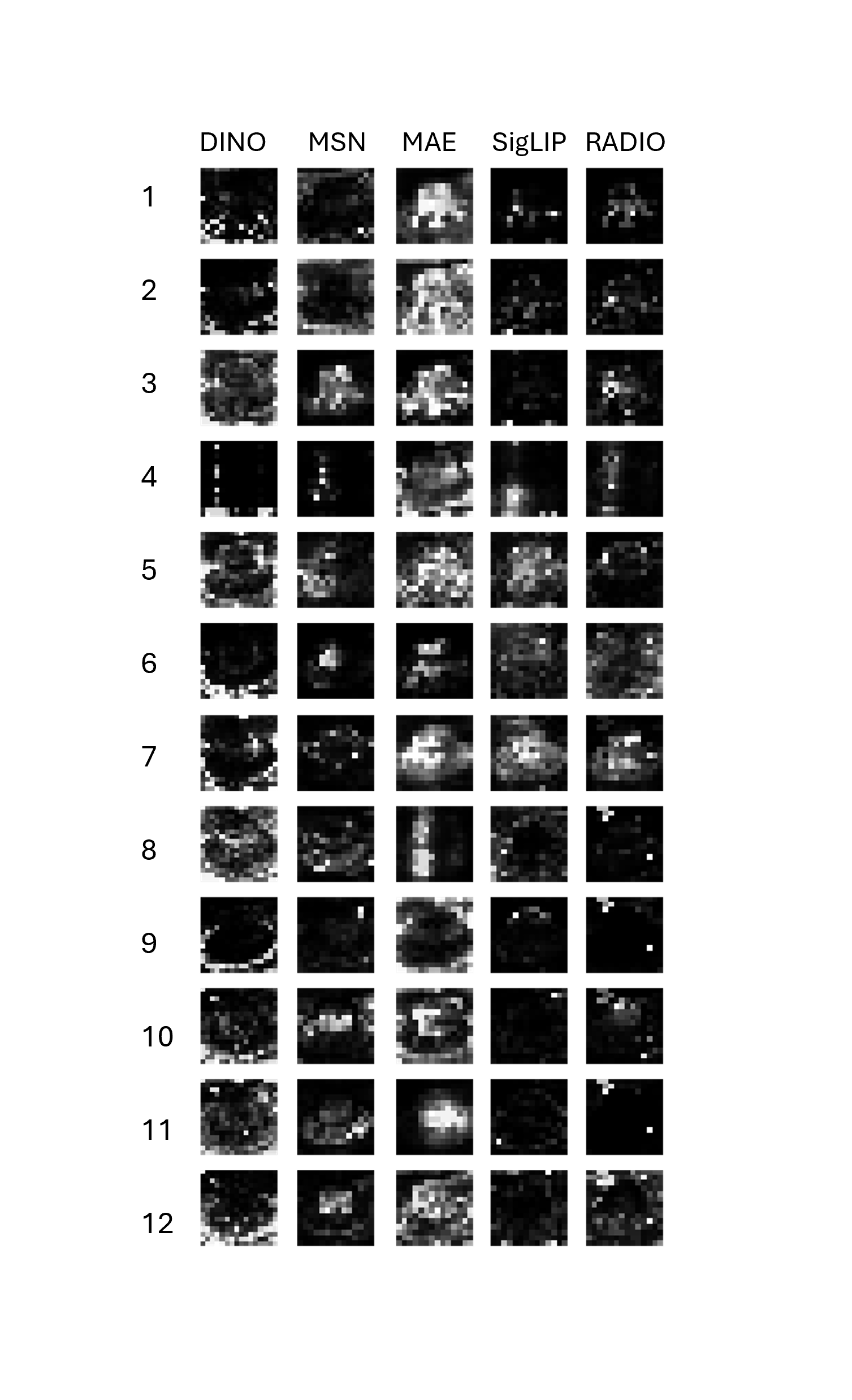}}
\caption{Attention maps for all models, MGCLS multi-source crop (left).  ImageNet (right), shown for each model and each attention block.} \label{fig:mgclsattention}
\end{figure*}

We visualize the attention maps for sample images from each dataset, which are shown in Figure \ref{fig:attentionims}. One map per attention block is shown; ViT-Base, the backbone architecture for all Vision Transformers used in this study, has twelve blocks. 

Maps of blocks with higher numbers, which typically capture high-level semantic relationships as opposed to earlier blocks which focus on more low-level, localized features, do not always display meaningful features when calculated for astrophysical images (Figures \ref{fig:rgzattention}, \ref{fig:gmnistattention} and \ref{fig:mgclsattention} left side). This is especially evident with SigLIP and AM-RADIO. Models pre-trained on natural images only (DINOv2, MSN, and MAE), however, do display the typical behavior described above in attention maps calculated for an ImageNet example (Figure \ref{fig:mgclsattention}, right side). Examination of the attention maps therefore illustrates that the semantic relationships, both local and global, learned from natural image pre-training do not always have meaningful counterparts in astrophysical images.


\end{appendix}

\end{document}